\documentclass[11pt,a4paper]{article}
\pdfoutput=1

\newcommand{\publishedID}{JCAP 06 (2014) 054}
\usepackage{jcappub}

\usepackage{mathrsfs}
\usepackage{bm}
\usepackage{dsfont}
\usepackage{tensor}
\usepackage{xspace}
\usepackage{lmodern}
\usepackage{wasysym}
\usepackage{microtype}

\newcommand{\dd}{{\rm d}}
\newcommand{\pd}[3][]{\frac{\partial^{#1} #2}{\partial {#3}^{#1}}}
\newcommand{\ddf}[3][]{\frac{\dd^{#1} #2}{\dd {#3}^{#1}}}
\newcommand{\ph}{\varphi}
\newcommand{\eps}{\varepsilon}
\newcommand{\define}{\equiv}
\newcommand{\mean}[1]{\left\langle #1 \right\rangle}

\newcommand{\pa}[1]{\left( #1 \right)}
\newcommand{\pac}[1]{\left[ #1 \right]}
\newcommand{\paac}[1]{\left\{ #1 \right\}}

\newcommand{\identity}{\boldsymbol{1}}
\newcommand{\zero}{\boldsymbol{0}}
\newcommand{\separation}{\boldsymbol{\mathcal{\xi}}}
\newcommand{\optical}{\boldsymbol{\mathcal{R}}}
\newcommand{\jacobi}{\boldsymbol{\mathcal{D}}}
\newcommand{\amplification}{\boldsymbol{\mathcal{A}}}
\newcommand{\scale}{\boldsymbol{\mathcal{C}}}
\newcommand{\wronski}{\boldsymbol{\mathcal{W}}}

\newcommand{\KK}{\tilde{K}}
\newcommand{\aK}{\tilde{a}}
\newcommand{\HK}{\tilde{H}}
\newcommand{\rhoK}{\tilde{\rho}}

\newenvironment{system}{\left\{\begin{aligned}}{\end{aligned}\right.}

\title{Swiss-cheese models and the Dyer-Roeder approximation}

\author{Pierre Fleury}

\affiliation{
Institut d'Astrophysique de Paris, UMR-7095 du CNRS, Universit\'e Pierre et Marie Curie,\\98 bis, boulevard Arago, 75014 Paris, France.\\
Sorbonne Universit\'es, Institut Lagrange de Paris,\\ 98 bis, boulevard Arago, 75014 Paris,~France.
}

\emailAdd{fleury@iap.fr}  

\abstract{In view of interpreting the cosmological observations precisely, especially when they involve narrow light beams, it is crucial to understand how light propagates in our statistically homogeneous, clumpy, Universe. Among the various approaches to tackle this issue, Swiss-cheese models propose an inhomogeneous space-time geometry which is an exact solution of Einstein's equation, while the Dyer-Roeder approximation deals with inhomogeneity in an effective way. In this article, we demonstrate that the distance-redshift relation of a certain class of Swiss-cheese models is the same as the one predicted by the Dyer-Roeder approach, at a well-controlled level of approximation. Both methods are therefore equivalent when applied to the interpretation of, e.g., supernova observations. The proof relies on completely analytical arguments, and is illustrated by numerical results.
}


\arxivnumber{1402.3123}

\date{\today}
\begin{document}

\maketitle

\flushbottom

%%%%%%%%%%%%%%%%%%%%%%%%%%%%%%%%%%%%%%%%%%%%%%%%%%%%%%%%%%%%%%%%
%%%%%%%%%%%%%%%%%%%%%%%%%%%%%%%%%%%%%%%%%%%%%%%%%%%%%%%%%%%%%%%%
\section*{Introduction}
\addcontentsline{toc}{section}{Introduction}
%%%%%%%%%%%%%%%%%%%%%%%%%%%%%%%%%%%%%%%%%%%%%%%%%%%%%%%%%%%%%%%%
%%%%%%%%%%%%%%%%%%%%%%%%%%%%%%%%%%%%%%%%%%%%%%%%%%%%%%%%%%%%%%%%

All cosmological observations involve, today, exclusively photons as the carrier of the information. In order to interpret them correctly, it is thus primordial to understand how light propagates through the Universe. In particular, the relation between the angular diameter distance~$D_{\rm A}$ (or the luminosity distance~$D_{\rm L}$) and the redshift~$z$ of remote sources, is a key ingredient both in the interpretation of the baryon acoustic oscillation (BAO) signal, whether it is extracted from the correlation function of the matter distribution~\cite{2013arXiv1312.4877A,2014arXiv1404.1801D} or from the anisotropies of the cosmic microwave background (CMB)~\cite{2013arXiv1303.5076P}; and, of course, in the analysis of the Hubble diagram, constructed from supernova (SN) observations~\cite{2011ApJS..192....1C,2012ApJ...746...85S}.

Though crucial, the determination of a reliable optical model of our Universe, known as the fitting problem~\cite{1987CQGra...4.1697E}, still remains to be done. In practice, observational cosmologists always rely on the somehow least worst model, in which light propagates through a Friedmann-Lema\^itre (FL) space-time, describing a perfectly homogeneous and isotropic universe~\cite{PeterUzan}. While such an approximation may be valid for wide light beams (e.g., involved in BAO observations), typically sensitive to the large-scale structure of the Universe, it is much more questionable regarding the very narrow beams involved in astronomical observations, e.g., SNe~\cite{2013PhRvL.111i1302F}.

Of course, the challenge of establishing a better optical model for the Universe led to many studies based on various methods. Popular ones, in the paradigm of standard cosmology, consist in the analysis of weak lensing in a perturbed FL space-time~\cite{2000A&A...354..767V,2006PhRvD..73b3523B,2011PhRvD..83b3009K,
2012PhRvD..86b3510D,2012arXiv1207.2109U,2012JCAP...11..045B,2013JCAP...06..002B,
2013arXiv1309.6542N,2014arXiv1402.1933U}, or in cosmological simulations~\cite{2009A&A...497..335T}. Alternative relativistic models for the inhomogeneous Universe can also be considered, such as Swiss-cheese models~\cite{1967ApJ...150..737G,
1969ApJ...155...89K,1974ApJ...189..167D,2007JCAP...02..013B,
2007PhRvD..76l3004M,2008JCAP...04..008B,2008JCAP...06..021B,
2008PhRvD..78h3511V,2009JCAP...06..010V,2009MNRAS.400.2185C,
2011PhRvD..84d4011S,2012PhRvD..85b3510F}, lattice models~\cite{1957RvMP...29..432L,2013CQGra..30b5002B}, or plane-symmetric models~\cite{2014arXiv1401.3634A}. Finally, rather than specifying any space-time model, one can use simplifying assumptions about the impact of the inhomogeneity of matter distribution on light propagation, in order derive an effective model. It is the case of the Dyer-Roeder approach~\cite{1972ApJ...174L.115D}, inspired from Zel'dovich's original intuition~\cite{1964AZh....41...19Z}. We refer the reader to, e.g., Refs.~\cite{2012JCAP...05..003B,2012MNRAS.426.1121C} for elements of review and comparison.

Among all those approaches, the Dyer-Roeder~(DR) approximation on the one hand, and the ``traditional'' Swiss-cheese~(SC) models generated by the Einstein-Straus method~\cite{1945RvMP...17..120E,1946RvMP...18..148E} on the other hand, used to be studied in parallel and presented together (see, e.g., Textbook~\cite{1992grle.book.....S}). This is actually not surprising, because, in its origin, the DR approximation was motivated by such SC models. Though very different in their philosophy---the former is an effective theory, based on assumptions, while the latter relies on a well-defined space-time model---, both approaches seem to generate similar distance-redshift relations~\cite{2013PhRvD..87l3526F}
\begin{equation}
D_{\rm A}^{\rm DR}(z)\approx D_{\rm A}^{\rm SC}(z).
\label{eq:correspondence}
\end{equation}
However, to the knowledge of the author, such a correspondence has never been explained, nor rigorously proved, in the literature. The purpose of this article is thus to fill the blank, not only by checking the \emph{conjecture} \eqref{eq:correspondence} numerically, but also by proposing an analytical proof of it, in order to understand the underlying mechanisms, and its domain of validity.

In Sec.~\ref{sec:geometric_optics}, we recall theoretical elements about geometric optics, needed for the remainder of the article. In Secs.~\ref{sec:DR} and \ref{sec:Swiss-cheese_models} we introduce, respectively, the DR approximation and SC models. Section~\ref{sec:optics_SC} is then dedicated to the analysis of the optical properties of SC models, that we prove to be equivalent to the ones predicted by the DR approach, at a very good level of approximation. Finally, in Sec.~\ref{sec:numerical_results}, we propose numerical illustrations of our results, and we analyse the origin of the small discrepancies between the SC and DR approaches.

%%%%%%%%%%%%%%%%%%%%%%%%%%%%%%%%%%%%%%%%%%%%%%%%%%%%%%%%%%%%%%%%
%%%%%%%%%%%%%%%%%%%%%%%%%%%%%%%%%%%%%%%%%%%%%%%%%%%%%%%%%%%%%%%%
\section{Geometric optics in curved space-time}
\label{sec:geometric_optics}
%%%%%%%%%%%%%%%%%%%%%%%%%%%%%%%%%%%%%%%%%%%%%%%%%%%%%%%%%%%%%%%%
%%%%%%%%%%%%%%%%%%%%%%%%%%%%%%%%%%%%%%%%%%%%%%%%%%%%%%%%%%%%%%%%

This section reviews some generic elements about the propagation of light in arbitrary space-times. We define our notations, and introduce several tools which will be useful in the remainder of the article.

%%%%%%%%%%%%%%%%%%%%%%%%%%%%%%%%%%%%%%%%%%%%%%%%%%%%%%%%%%%%%%%%
\subsection{Description of a light beam}
%%%%%%%%%%%%%%%%%%%%%%%%%%%%%%%%%%%%%%%%%%%%%%%%%%%%%%%%%%%%%%%%

A light beam is a collection of light rays, that is, a bundle of null geodesics $\{x^\mu(v,r)\}$, where $r$ labels the rays and $v$ is the affine parameter along them. The wave four-vector $k^\mu \define \partial x^\mu/\partial v$ is a null vector field, tangent to the rays $r=\mathrm{cst}$. It therefore satisfies
\begin{equation}
k^\mu k_\mu = 0, \qquad k^\nu \nabla_\nu k_\mu = 0.
\end{equation}

Besides, the relative behaviour of two neighbouring rays $x^\mu(\cdot,r)$ and $x^\mu(\cdot,r+\dd r)$ is described by their separation vector $\xi^\mu \define \partial x^\mu/\partial r$. One can always choose the origin of the affine parametrization of each ray $r=\mathrm{cst}$ so that
\begin{equation}
k^\mu \xi_\mu = 0.
\label{eq:orthogonality_separation_wave}
\end{equation}
Note that this condition is automatically satisfied if one sets $v=0$, for each geodesic, at a vertex point of the bundle, that is an event where $\xi^\mu=0$. When the condition~\eqref{eq:orthogonality_separation_wave} is satisfied, the evolution of $\xi^\mu$ along the light beam is governed by the geodesic deviation equation
\begin{equation}
k^\alpha k^\beta \nabla_\alpha \nabla_\beta \xi^\mu = {R^\mu}_{\nu\alpha\beta} k^\nu k^\alpha\xi^\beta,
\label{GDE}
\end{equation}
where ${R^\mu}_{\nu\alpha\beta}$ is the Riemann tensor.

%%%%%%%%%%%%%%%%%%%%%%%%%%%%%%%%%%%%%%%%%%%%%%%%%%%%%%%%%%%%%%%%
\subsection{The Sachs formalism}
%%%%%%%%%%%%%%%%%%%%%%%%%%%%%%%%%%%%%%%%%%%%%%%%%%%%%%%%%%%%%%%%

Consider an observer, with four-velocity $u^\mu$ ($u_\mu u^\mu=-1$), who crosses the light beam. With respect to this observer, one defines the spatial direction of light propagation as the opposite of the only direction for which the observer can detect a signal. It is spanned by the purely spatial unit vector $d^\mu$,
\begin{equation}
d^\mu u_\mu = 0, \qquad d^\mu d_\mu = 1,
\end{equation}
which leads to the 3+1 decomposition of the wave four-vector
\begin{equation}
k^\mu =  \omega (u^\mu - d^\mu),
\end{equation}
where $\omega = 2\pi \nu \define -u_\mu k^\mu$ is the cyclic frequency of the light signal in the observer's rest frame. Note that $\dd \ell = \omega \dd v$ is the proper distance (measured by the observer) travelled by light for a change $\dd v$ of the affine parameter. The redshift $z$ is defined as the relative difference between the emitted frequency~$\nu_{\rm s}$, in the source's frame, and the observed frequency~$\nu_{\rm o}$, in the observer's frame, so that
\begin{equation}
1+z \define \frac{\nu_{\rm s}}{\nu_{\rm o}} 
= \frac{u_{\rm s}^\mu k_\mu(v_{\rm s})}{u_{\rm o}^\mu k_\mu(v_{\rm o})} .
\end{equation}

Now suppose that the observer wishes to measure the size and the shape of the light beam. For that purpose, he must use a (spatial) screen orthogonal to the line of sight. This screen is spanned by the so-called Sachs basis $(s_A^\mu)_{A\in\{1,2\}}$, defined by
\begin{equation}
s_{A}^\mu u_{\mu}=s_{A}^\mu d_{\mu}=0,
\qquad
g_{\mu\nu }s_{A}^\mu s_{B}^ \nu=\delta_{AB},
\end{equation}
and by the transport property \eqref{eq:transport_Sachs} below. The projections $\xi_A \define s_A^\mu \xi_\mu$ indicate the relative position, on the observer's screen, of the light points corresponding to two neighbouring rays separated by $\xi^\mu$. Thus, it encodes all the information about the size and shape of the beam.

Consider a family of observers $u^\mu(v)$, along the beam, who wants to follow the evolution of the shape of the beam (typically for shear measurements). For that purpose, they must all use the ``same'' Sachs basis, in order to avoid any spurious rotation of the pattern observed on the  screens. This is ensured by a partial parallel transportation
\begin{equation}
S_{\mu\nu} k^\rho \nabla_\rho s_A^\nu = 0 ,
\label{eq:transport_Sachs}
\end{equation}
where $S^{\mu\nu} = \delta^{AB} s_A^\mu s_B^\nu = g\indices{^\mu^\nu}+u^\mu u^\nu - d^\mu d^\nu$ is the screen projector. The reason why $s_A^\mu$ cannot be completely parallel-transported is that, in general, $u^\mu$ is not\footnote{In fact, it is also possible to choose a family of observers such that the four-velocity field $u^\mu$ is parallel-transported along the beam, without affecting the optical equations~\cite{1992grle.book.....S}. In this case, however, the observers are generally not comoving, and thus have no clear cosmological interpretation.}.

The evolution of $\xi_A$, with light propagation, is determined by projecting the geodesic deviation equation \eqref{GDE} on the Sachs basis. The result is known as the Sachs equation~\cite{1961RSPSA.264..309S,1992grle.book.....S},
\begin{equation}
\frac{\dd^2\xi_A}{\dd v^2} = \mathcal{R}_{AB} \, \xi^B,
\label{eq:Sachs}
\end{equation}
where $\mathcal{R}_{AB}={R}_{\mu\nu\alpha\beta}k^\nu k^\alpha s_A^\mu s_B^\beta$ is the screen-projected Riemann tensor, called the optical tidal matrix. The properties of the Riemann tensor imply that this matrix is symmetric, $\mathcal{R}_{AB}=\mathcal{R}_{BA}$. Note that the altitude of the ``screen indices'' ($A, B, \ldots$) does not matter, since they are raised and lowered by $\delta_{AB}$. In the following, to alleviate the notation, we use bold symbols for quantities with screen indices, and an overdot for derivatives with respect to the affine parameter~$v$. The Sachs equation~\eqref{eq:Sachs} thus becomes $\ddot{\separation}=\optical \separation$.

The Riemann tensor can be decomposed into a Ricci part and a Weyl part,
\begin{equation}
R\indices{_\mu_\nu_\alpha_\beta} = 
g\indices{_\mu_[_\alpha} R\indices{_\beta_]_\nu} 
- g\indices{_\nu_[_\alpha} R\indices{_\beta_]_\mu}
- \frac{1}{3} R\,g\indices{_\mu_[_\alpha} g\indices{_\beta_]_\nu}
+ C\indices{_\mu_\nu_\alpha_\beta} ,
\end{equation}
where the Ricci tensor $R_{\mu\nu}$ is directly related to the local density of energy-momentum via Einstein's equations; and the Weyl tensor $C_{\mu\nu\alpha\beta}$ contains the long-range effects of gravitation. As a consequence, the optical tidal matrix can also be split into a pure-trace Ricci-lensing term and a traceless Weyl-lensing term as
\begin{equation}\label{eq:decomposition_optical_matrix}
\optical =
\underbrace{
\begin{pmatrix}
\Phi_{00} & 0 \\ 0 & \Phi_{00}
\end{pmatrix}
}_{\text{Ricci lensing}}
+
\underbrace{
\begin{pmatrix}
- {\rm Re}\,\Psi_0 & {\rm Im}\,\Psi_0 \\
  {\rm Im}\,\Psi_0 & {\rm Re}\,\Psi_0
\end{pmatrix}
}_{\text{Weyl lensing}},
\end{equation}
with
\begin{equation}\label{eq:Ricci-Weyl}
\Phi_{00} \define -\frac12 R_{\mu\nu} k^\mu k^\nu,
\qquad \text{and} \qquad
\Psi_0 \define -\frac{1}{2} C_{\mu\nu\alpha\beta}
(s_1^{\mu} - {\rm i} s_2^{\mu}) k^\nu k^\alpha (s_1^{\beta} - {\rm i} s_2^{\beta}).
\end{equation}
It is then clear, from the Sachs equation~\eqref{eq:Sachs}, that the Ricci term tends to isotropically focus the light beam, while the Weyl term tends to shear it. For this reason, $\Phi_{00}$ is called ``source of convergence'' and $\Psi_{0}$ ``source of shear\footnote{This name, however, omits a part of the optical effects due to $\Psi_0$; strictly speaking, we should write ``source of shear \emph{and rotation}''. Indeed, even though the beam is an irrotational bundle of null geodesics ($\nabla_{[\mu}k_{\nu]}=0$), a rotation of the image can appear due to cumulative shearing along different directions.}'' \cite{2001PhR...340..291B}.

%%%%%%%%%%%%%%%%%%%%%%%%%%%%%%%%%%%%%%%%%%%%%%%%%%%%%%%%%%%%%%%%
\subsection{Wronski matrix, Jacobi matrix}
\label{subsec:wronski_matrix}
%%%%%%%%%%%%%%%%%%%%%%%%%%%%%%%%%%%%%%%%%%%%%%%%%%%%%%%%%%%%%%%%

Because the Sachs equation is a second-order homogeneous linear differential equation, any solution is linearly related to its initial conditions ($v=v_0$), so that
\begin{align}
\separation(v) &= \scale(v \leftarrow v_0) \separation(v_0)
+ \jacobi(v \leftarrow v_0) \dot{\separation}(v_0),
\label{eq:separation}\\
\dot{\separation}(v) &= \dot{\scale}(v \leftarrow v_0) \separation(v_0)
			+ \dot{\jacobi}(v \leftarrow v_0) \dot{\separation}(v_0),
\label{eq:separation_dot}
\end{align}
where $\scale(v \leftarrow v_0)$ and $\jacobi(v \leftarrow v_0)$ are $2 \times 2$ matrices, respectively called scale matrix and Jacobi matrix, which satisfy the Sachs equation like $\separation(v)$, with initial conditions
\begin{equation}
\begin{cases}
\scale(v_0 \leftarrow v_0) = \identity_2 \\
\dot{\scale}(v_0 \leftarrow v_0) = \zero_2
\end{cases}
\qquad \mathrm{and} \qquad
\begin{cases}
\jacobi(v_0 \leftarrow v_0) = \zero_2 \\
\dot{\jacobi}(v_0 \leftarrow v_0) = \identity_2
\end{cases} ,
\end{equation}
where $\zero_n$ and $\identity_n$ denote respectively the $n\times n$ zero and identity matrices. Equations \eqref{eq:separation}, \eqref{eq:separation_dot} can finally be gathered into a single $4\times 4$ matrix relation:
\begin{equation}
\begin{pmatrix}
\separation\\
\dot{\separation}
\end{pmatrix}(v)
=
\wronski(v \leftarrow v_0)
\begin{pmatrix}
\separation\\
\dot{\separation}
\end{pmatrix}(v_0),
\qquad
\mathrm{where}
\qquad
\wronski \define
\begin{pmatrix}
\scale & \jacobi \\
\dot{\scale} & \dot{\jacobi}
\end{pmatrix}
\end{equation}
is the $4 \times 4$ Wronski matrix of the Sachs equation. 
%In the context of gravitational lensing, this tool was first introduced in Ref.~\cite{2013PhRvD..87l3526F}.
As we will see in Sec. \ref{sec:optics_SC}, it is particularly convenient for dealing with light propagation through a patchwork of space-times, such as Swiss-cheese models, because by construction
\begin{equation}\label{eq:Chasles_Wronski}
\wronski(v_3 \leftarrow v_1) = \wronski(v_3 \leftarrow v_2) \wronski(v_2 \leftarrow v_1).
\end{equation}
It is easy to see that the Wronski matrix is the only solution of
\begin{equation}\label{eq:master_equation_Wronski}
\dot{\wronski}(v \leftarrow v_0) =
\begin{pmatrix}
\zero_2 & \identity_2 \\
\optical(v) & \zero_2
\end{pmatrix}
\wronski(v \leftarrow v_0)
\qquad \mathrm{with} \qquad \wronski(v_0 \leftarrow v_0) = \identity_4 .
\end{equation}
This differential equation is formally solved by
\begin{equation}\label{eq:formal_solution_Wronski}
\wronski(v \leftarrow v_0)
= \mathrm{Vexp} \int_{v_0}^{v}
\begin{pmatrix}
\zero_2 & \identity_2 \\
\optical(w) & \zero_2
\end{pmatrix} \dd w ,
\end{equation}
where $\mathrm{Vexp}$ is the affine-parameter ordered exponential, analogous to the time-ordered exponential in quantum field theory. It is defined, for any matrix-valued function $\boldsymbol{\mathcal{M}}$, by
\begin{equation}
\mathrm{Vexp}\int_{v_0}^{v} \boldsymbol{\mathcal{M}}(w) \dd w
\define
\sum_{n=0}^{\infty} \int_{v_0}^v \dd w_1 \int_{v_0}^{w_1} \dd w_2 \ldots \int_{v_0}^{w_{n-1}} \dd w_n \;
					\boldsymbol{\mathcal{M}}(w_1) \boldsymbol{\mathcal{M}}(w_2) \ldots \boldsymbol{\mathcal{M}}(w_n) .
\end{equation}
This expression reduces to a regular exponential if, for all $v$, $v'$, $\boldsymbol{\mathcal{M}}(v)$ commutes with $\boldsymbol{\mathcal{M}}(v')$. In the case of Eq.~\eqref{eq:formal_solution_Wronski}, this apples if, and only if, the optical tidal matrix $\optical(v)$ is a constant. 

%%%%%%%%%%%%%%%%%%%%%%%%%%%%%%%%%%%%%%%%%%%%%%%%%%%%%%%%%%%%%%%%
\subsection{Angular distance and luminosity distance}
%%%%%%%%%%%%%%%%%%%%%%%%%%%%%%%%%%%%%%%%%%%%%%%%%%%%%%%%%%%%%%%%

The observational notion of angular distance $D_{\rm A}$, which relates the emission cross-sectional area $\dd^2 A_{\rm s}$ of a source to the observed angular aperture $\dd \Omega_{\rm o}^2$, via
\begin{equation}
\dd^2 A_{\rm s} = D_{\rm A}^2 \dd \Omega_{\rm o}^2,
\end{equation}
is naturally related to the Jacobi part $\jacobi$ of the Wronski matrix. Indeed, on the one hand $\separation(v_{\rm s})=\boldsymbol{\ell}_{\rm s}$ is the proper separation (in the source's frame) between two emission points within the extended source; on the other hand $\dot{\separation}(v_{\rm o})/\omega_{\rm o}=\boldsymbol{\theta}_{\rm o}$ is the observed angular separation between the light rays emitted by these points. Thus, from Eq.~\eqref{eq:separation}, we find
\begin{equation}
\omega_{\rm o}\jacobi(v_{\rm s} \leftarrow v_{\rm o})
= \frac{\partial\separation(v_{\rm s})}
		{\partial\dot{\separation}(v_{\rm o})/\omega_{\rm o}}
= \frac{\partial\boldsymbol{\ell}_{\rm s}}
		{\partial \boldsymbol{\theta}_{\rm o}},
\end{equation}
so that
\begin{equation}
D_{\rm A} = \sqrt{\det \omega_{\rm o} \jacobi(\mathrm{s}\leftarrow \mathrm{o})}.
\end{equation}
The observational luminosity distance~$D_{\rm L}$, relating the source's intrinsic luminosity~$L_{\rm s}$ and the observed flux~$F_{\rm o}$ via $L_{\rm s}=4\pi D_{\rm L}^2 F_{\rm o}$, can also be expressed in terms of $\jacobi$~\cite{2004LRR.....7....9P} according to
\begin{equation}
D_{\rm L}
=(1+z)\sqrt{\det \omega_{\rm s} \jacobi(\mathrm{o}\leftarrow \mathrm{s})}
=(1+z)^2 D_{\rm A}.
\label{eq:duality_law}
\end{equation}
We stress that, contrary to what it is sometimes wrongly believed, the duality law \eqref{eq:duality_law} is true for \emph{any} space-time, as far as the number of photons is conserved during light travel.

Since the Jacobi matrix $\jacobi$ not only encodes information about the size of the beam, but also about its shape, all the weak-lensing observational quantities (convergence, shear, magnification) can be extracted from it; see, e.g., Ref.~\cite{2001PhR...340..291B} for more details. Moreover, some genuinely relativistic effects, such as optical rotation, which are usually not taken into account by weak lensing studies, are also encoded in $\jacobi$; Ref.~\cite{2013PhRvD..87d3003P} provides an example in the context of anisotropic cosmology. Let us finally indicate that, by a suitable choice of coordinates adapted to the lightcone, called GLC gauge~\cite{2011JCAP...07..008G} (inspired from the observational coordinates~\cite{1985PhR...124..315E}), the expression of the Jacobi matrix can be trivialized~\cite{2013JCAP...11..019F}, so that the whole information is, in this case, contained in the Sachs basis only.

%%%%%%%%%%%%%%%%%%%%%%%%%%%%%%%%%%%%%%%%%%%%%%%%%%%%%%%%%%%%%%%%
%%%%%%%%%%%%%%%%%%%%%%%%%%%%%%%%%%%%%%%%%%%%%%%%%%%%%%%%%%%%%%%%
\section{The Dyer-Roeder approximation}
\label{sec:DR}
%%%%%%%%%%%%%%%%%%%%%%%%%%%%%%%%%%%%%%%%%%%%%%%%%%%%%%%%%%%%%%%%
%%%%%%%%%%%%%%%%%%%%%%%%%%%%%%%%%%%%%%%%%%%%%%%%%%%%%%%%%%%%%%%%

In this section, we describe in detail the propagation of light in a homogeneous and isotropic universe, and how it must be modified according to the Dyer-Roeder (DR) prescription. The last subsection is dedicated to a discussion about its physical motivations and its limitations.

%%%%%%%%%%%%%%%%%%%%%%%%%%%%%%%%%%%%%%%%%%%%%%%%%%%%%%%%%%%%%%%%
\subsection{Light propagation in a homogeneous and isotropic universe}
%%%%%%%%%%%%%%%%%%%%%%%%%%%%%%%%%%%%%%%%%%%%%%%%%%%%%%%%%%%%%%%%

Let us apply the formalism developed in the previous section to the Friedmann-Lema\^itre~(FL) geometry. The associated metric reads (in three different coordinate systems)
\begin{align}
\dd s^2 &= -\dd T^2 + a^2(T) \pac{\frac{\dd R^2}{1-K R^2} + R^2 \dd\Omega^2}\label{eq:FL_metric}\\
		&= -\dd T^2 + a^2(T) \pac{\dd\chi^2 + f_K(\chi)^2 \dd\Omega^2} \\
		&= a^2(\eta) \pac{-\dd\eta^2 + \dd\chi^2 + f_K(\chi)^2 \dd\Omega^2},
\end{align}
where $\dd\Omega^2 \define \dd\theta^2+\sin^2\theta \dd\ph^2$ is the infinitesimal solid angle; $T$, $\eta$ denote respectively the cosmic and conformal times, with $\dd T=a \dd\eta$; $a$ is the scale factor; $\chi$ is the comoving radius, $R=f_K(\chi)$ the comoving areal radius, with
\begin{equation}
f_K(\chi) \define
\begin{system}
&\sin(\sqrt{K}\chi)/\sqrt{K} & \quad & \mbox{if } K>0 \\
&\chi & \quad & \mbox{if } K=0 \\
&\sinh(\sqrt{-K}\chi)/\sqrt{-K} & \quad & \mbox{if } K<0
\end{system};
\end{equation}
and finally $6 K/a^2$ is the (intrinsic) scalar curvature of the $T=\mathrm{cst}$ spatial hyper-surfaces. The time evolution of the scale factor $a(T)$ is ruled by the Friedmann equation
\begin{equation}\label{eq:Friedmann}
H^2 \define \pa{ \frac{1}{a} \ddf{a}{T} }^2 = \frac{8\pi G \rho_0}{3} \pa{\frac{a_0}{a}}^3 - \frac{K}{a^2} + \frac{\Lambda}{3},
\end{equation}
where $\rho$ is the homogeneous energy density of matter, modelled by a dust fluid filling space, and $\Lambda$ is the cosmological constant. As usual, a subscript $0$ denotes the present value of a quantity. The Friedmann equation can be also written in terms of the cosmological parameters~$\{\Omega\}$,
\begin{equation}
H^2 = H_0^2 \pac{ \Omega_{\rm m0} \pa{\frac{a_0}{a}}^3 + \Omega_{K0} \pa{\frac{a_0}{a}}^2 + \Omega_{\rm \Lambda 0} },
\end{equation}
with
\begin{equation}
\Omega_{\rm m0} \define \frac{8\pi G \rho_0}{3 H_0^2},
\qquad
\Omega_{K0} \define \frac{-K}{a_0^2 H_0^2},
\qquad
\Omega_{\Lambda 0} \define \frac{\Lambda}{3 H_0^2}.
\end{equation}

We now focus on light propagation. Consider a comoving observer, who can be chosen without loss of generality at the origin of the spatial coordinate system. A light ray reaching this central observer today is purely radial, and propagates according to $\chi=\eta_0-\eta$. Along it, the affine parameter $v$ satisfies $\dd\eta/\dd v = \omega/a$, and $a\omega$ is a constant (whence the FL expression for the redshift, $1+z=a_{\rm o}/a_{\rm s}$). The evolution of the redshift with the affine parameter is therefore ruled by
\begin{equation}\label{eq:v_z_FL}
%\left.
\frac{1}{\omega_{\rm o}}\ddf{}{v}\pa{ \frac{1}{1+z} }
%\right|_{\rm FL}
= H.
\end{equation}

The screen vectors $s_1$, $s_2$, forming the Sachs basis, do not need here to be specified explicitly to get the optical tidal matrix $\optical$, because of the high degree of symmetry (in particular, spatial isotropy) of the FL space-time. The result is
\begin{equation}\label{eq:optical_FL}
\optical_{\rm FL} = -4 \pi G \rho \omega^2 \identity_2 .
\end{equation}
As expected, a FL space-time only focusses light via a Ricci term (source of convergence), because conformal flatness imposes that the Weyl tensor (source of shear and rotation) vanishes. The Sachs equation~\eqref{eq:Sachs} can then be solved exactly, e.g., by taking advantage of the conformal flatness \citep{2013PhRvD..87l3526F}, in order to obtain the blocks of the Wronski matrix:
\begin{align}
\scale_{\rm FL}(2 \leftarrow 1) &= \frac{a_2}{a_1} 
\Big[
f'_K(\eta_2 - \eta_1) - \mathcal{H}_1\,f_K(\eta_2-\eta_1) \Big] \identity_2, 
\label{eq:scale_FL}
\\[1mm]
\dot{\scale}_{\rm FL}(2 \leftarrow 1) &= \frac{\omega_2}{a_2}
\paac{
\mathcal{H}_2 \, \scale(2 \leftarrow 1) - \frac{a_2}{a_1} 
\Big[ 
K\,f_K(\eta_2-\eta_1) + \mathcal{H}_1\,f'_K(\eta_2-\eta_1)
\Big] \identity_2
},
\label{eq:scaledot_FL}
\\[3mm]					
\omega_1 \jacobi_{\rm FL}(2 \leftarrow 1) &=a_2 \, f_K(\eta_2-\eta_1) \, \identity_2,
\label{eq:jacobi_FL}
\\[2mm]
\omega_1 \dot{\jacobi}_{\rm FL}(2 \leftarrow 1) &= \frac{a_1}{a_2^2}\,\mathcal{H}_2\,[\omega_1 \jacobi(2 \leftarrow 1)] + \omega_1 \, \frac{a_1}{a_2} \, f'_K(\eta_2-\eta_1) \, \identity_2,
\label{eq:jacobidot_FL}
\end{align}
where $\mathcal{H} \define a'(\eta)/a = aH$ is the conformal Hubble parameter, and a prime denotes a derivative with respect to conformal time $\eta$. Note that \eqref{eq:jacobi_FL} gives the well-known expression for the angular distance in a FL universe,
\begin{equation}\label{eq:DA_FL}
D_{\rm A}^{\rm FL} 
= \sqrt{\det\omega_{\rm o} \jacobi_{\rm FL}({\rm s \leftarrow o})}
= a_{\rm s} f_K(\chi_{\rm s}).
\end{equation}

Although it is not obvious when written under this form, \emph{Eq.~\eqref{eq:DA_FL} must be considered a relation between the angular distance and the affine parameter}, because it results from solving $\dd^2\jacobi/\dd v^2=\optical\jacobi$. From this point of view, the usual distance-redshift relation $D_{\rm A}(z)$---as used, e.g., for interpreting the SN data---arises from \emph{both} Eq.~\eqref{eq:v_z_FL} and Eq.~\eqref{eq:DA_FL}. The importance of such a remark will become clearer in the next subsection.

%%%%%%%%%%%%%%%%%%%%%%%%%%%%%%%%%%%%%%%%%%%%%%%%%%%%%%%%%%%%%%%%
\subsection{The Dyer-Roeder approximation}
%%%%%%%%%%%%%%%%%%%%%%%%%%%%%%%%%%%%%%%%%%%%%%%%%%%%%%%%%%%%%%%%

As first pointed out by Zel'dovich \cite{1964AZh....41...19Z}, at the scale of the typical cross-sectional area of a light beam involved in astronomical observations, such as SNe, our Universe cannot be reasonably considered as homogeneously filled by a fluid, but rather composed of more or less concentrated clumps of matter. Therefore, the light signals involved in these observations must essentially propagate through vacuum, and consequently undergo focussing effects which are different from the FL case, presented in the previous subsection.

Such an intuition led Zel'dovich to propose an ``empty-beam'' approximation, generalized later  into a ``partially-filled-beam'' approach \cite{1965AZh....42..863D,1972ApJ...174L.115D}, better known today as the DR approximation. The aim is to provide an \emph{effective} distance-redshift relation $D_{\rm A}(z)$ which would take the small-scale inhomogeneity (i.e. the clumpiness) of our Universe into account. Such a relation can then be used for interpreting the SN data, instead of the standard FL one.

The DR approximation is based on three hypotheses:
\begin{description}
\item[DR1] The relation between the redshift $z$ and the affine parameter $v$ is essentially unaffected by the inhomogeneity of the distribution of matter.
\item[DR2] Weyl focussing is negligible regarding the evolution of the angular distance.
\item[DR3] Ricci focussing is effectively reduced, with respect to the FL case, by a factor $0\leq\alpha\leq 1$, called smoothness parameter, due to the fact that light mostly propagates through underdense regions of the universe. The physical meaning of $\alpha$ is thus the effective fraction of diffuse matter intercepted by the light beam during its propagation.
\end{description}
Those conditions imply that the DR relation between angular distance and redshift, $D_{\rm A}^{\rm DR}(z)$, is generated by solving both
\begin{align}
\frac{1}{\omega_{\rm o}}\ddf{}{v}\pa{ \frac{1}{1+z} } 
= H
\qquad &\text{(unchanged w.r.t. the FL case),}
\label{eq:z(v)_DR}
\\
\ddf[2]{\jacobi_{\rm DR}}{v} 
= \alpha \optical_{\rm FL} \jacobi_{\rm DR}
\qquad &\text{(reduced Ricci focussing, no Weyl focussing).}
\label{eq:Sachs_DR}
\end{align}
Note that, since $\optical_{\rm FL}\propto\identity_2$, the Jacobi matrix $\jacobi_{\rm DR}$ can be replaced in Eq.~\eqref{eq:Sachs_DR} by the square-root of its determinant, that is $D_{\rm A}^{\rm DR}$. Equations \eqref{eq:z(v)_DR} and \eqref{eq:Sachs_DR} can also be gathered in order to get a unique, second-order, differential equation
\begin{equation}\label{eq:DR_equation}
\ddf[2]{D_{\rm A}^{\rm DR}}{z} 
+ \pa{ \frac{2}{1+z} + \ddf{\ln H}{z} } \ddf{D_{\rm A}^{\rm DR}}{z}
+ \frac{3\alpha\Omega_{\rm m0}}{2} \pac{\frac{H_0}{H(z)}}^2 (1+z) D_{\rm A}^{\rm DR}(z)=0,
\end{equation}
known as the DR equation. In the original formulation of the DR approximation, the smoothness parameter~$\alpha$ was assumed to be a constant. However, according to its very definition, one can expect $\alpha$ (i) to depend on the line of sight, and (ii) to vary even along a given line of sight. In particular, it has been shown empirically~\cite{2011MNRAS.412.1937B} that, at least in a particular model for matter distribution, the DR equation gives results in good agreement with weak lensing if $\alpha-1\propto(1+z)^{-5/4}$. See also Ref.~\cite{2012MNRAS.426L..41B} for a discussion about how to measure $\alpha$ and test the DR approximation.

%%%%%%%%%%%%%%%%%%%%%%%%%%%%%%%%%%%%%%%%%%%%%%%%%%%%%%%%%%%%%%%%
\subsection{On the physical relevance of the approximation}
\label{subsec:relevance_DR}
%%%%%%%%%%%%%%%%%%%%%%%%%%%%%%%%%%%%%%%%%%%%%%%%%%%%%%%%%%%%%%%%

The physical relevance and the mathematical consistency of the DR approximation have been both questioned in the litterature \cite{1986A&A...168...57E,1989PhRvD..40.2502F,1993PThPh..90..753S,2009JCAP...02..011R,
2012MNRAS.426.1121C}. One of the criticisms, which lead to a ``modified DR approximation''~\cite{2012MNRAS.426.1121C,2013JCAP...11..020B}, relies on the argument that it is inconsistent to consider the Universe effectively underdense \emph{only} in the focussing term, and not in the $z(v)$ relation. In other words, hypotheses \textbf{DR1} and \textbf{DR3} would be incompatible.

In reaction to this argument, we stress that the essence of the DR approximation is precisely to notice that $z(v)$ and $D_{\rm A}(v)$ are ruled by the properties of the Universe considered \emph{at distinct scales}. On the one hand, $z(v)$ essentially\footnote{i.e., neglecting purely gravitational effects such as the (integrated) Sachs-Wolfe or Rees-Sciama effects.} depends on how the source and the observer move with respect to each other (adopting a Doppler-like interpretation of the cosmological redshift~\cite{2009AmJPh..77..688B}). he geodesic deviation equation indicates that this relative motion is governed by space-time curvature on the scale of the distance between the source and the observer. On the other hand, $D_{\rm A}(v)$ depends on the relative motion of two neighbouring rays within the beam, governed by space-time curvature on the scale of the beam itself. The ratio between both scales is given by the angular aperture of the beam, which is typically $\sim 10^{-10}$ for SN observations. Therefore, it is not inconsistent to suppose that a typical light beam could ``feel'' an underdense universe while the source and the observer do not. 

Let us close this section by a word on backreaction. It is known since the late $90$s that inhomogeneities of the distribution of matter in the Universe potentially affect its expansion averaged on cosmological scales (see, e.g., Refs.~\cite{2011arXiv1109.2314C,2011CQGra..28p4007B,
2013arXiv1311.3787W} for reviews). For the purpose of tracking such an effect in cosmological observations, one must wonder which properties of light propagation would be the most affected. Proceeding the rationale of the above paragraph, we expect the $D_{\rm A}(v)$ relation to be unaffected by any backreaction effect, because it involves too small scales. On the contrary, since the $z(v)$ relation has much more to do with a notion of global expansion, backreaction should have an impact on it. Therefore, one way of reading hypothesis \textbf{DR1} of the DR approximation is that it describes a clumpy universe with \emph{no backreaction}. This is, precisely, one of the main properties of the Swiss-cheese models presented in the next section (although this can be discussed, see Sec.~\ref{subsec:backreaction_SC}.)

%%%%%%%%%%%%%%%%%%%%%%%%%%%%%%%%%%%%%%%%%%%%%%%%%%%%%%%%%%%%%%%%
%%%%%%%%%%%%%%%%%%%%%%%%%%%%%%%%%%%%%%%%%%%%%%%%%%%%%%%%%%%%%%%%
\section{Swiss-cheese models}
\label{sec:Swiss-cheese_models}
%%%%%%%%%%%%%%%%%%%%%%%%%%%%%%%%%%%%%%%%%%%%%%%%%%%%%%%%%%%%%%%%
%%%%%%%%%%%%%%%%%%%%%%%%%%%%%%%%%%%%%%%%%%%%%%%%%%%%%%%%%%%%%%%%

\begin{figure}[!h]
\centering
\includegraphics[scale=1]{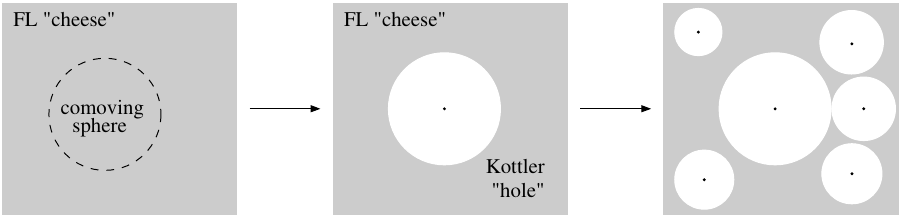}
\caption{Schematic construction of a Swiss-cheese model.}
\label{fig:SC_construction}
\end{figure}

Historically, Swiss-cheese~(SC) models were introduced by Einstein and Straus~\cite{1945RvMP...17..120E,1946RvMP...18..148E}, in 1945, as a method to embed a compact object within the expanding universe. It consists in removing a spherical comoving region from a FL space-time, and replacing it by a point mass at the center of the region (see Fig.~\ref{fig:SC_construction}). This creates a ``hole'' within the Friedmannian ``cheese'', and the operation can be repeated anywhere else, as long as the holes do not overlap. The reason why such a construction is possible is that the Schwarzschild (or Kottler) and FL geometries glue perfectly on a spherical frontier. This property can be justified (see, e.g., Refs.~\cite{1969ApJ...155...89K,2013PhRvD..87l3526F}) invoking the Darmois-Israel junction conditions~\cite{Darmois,1966NCimB..44....1I,1967NCimB..48..463I} between two space-times. In this section, we propose a slightly more intuitive approach.

%%%%%%%%%%%%%%%%%%%%%%%%%%%%%%%%%%%%%%%%%%%%%%%%%%%%%%%%%%%%%%%%
\subsection{Free-fall coordinates for the Kottler metric}
\label{subsec:free-fall_Kottler}
%%%%%%%%%%%%%%%%%%%%%%%%%%%%%%%%%%%%%%%%%%%%%%%%%%%%%%%%%%%%%%%%

The Kottler geometry~\cite{1918AnP...361..401K} is the extension of the Schwarzschild geometry to the case of a non-vanishing cosmological constant. Written with the usual Droste-Schwarzschild coordinates, the associated metric reads
\begin{equation}\label{eq:Kottler_metric}
\dd s^2 = -A(r)\,\dd t^2 + \frac{\dd r^2}{A(r)} + r^2 \dd\Omega^2 ,
\end{equation}
with $A(r) \define 1 - r_{\rm S}/r - \Lambda r^2/3$, $r_{\rm S}\define 2GM$ being the Schwarzschild radius associated with the central mass~$M$. It is possible to make this metric resemble the FL one \eqref{eq:FL_metric}, by using comoving and synchronous coordinates adapted to radially free-falling observers, analogous to the ones used to describe the Lema\^itre-Tolman-Bondi (LTB) geometry~\cite{1933ASSB...53...51L,1934PNAS...20..169T,
1947MNRAS.107..410B}. The construction is the following. Consider a test particle, which starts (at $t=0$) a radial free fall from $r=R$. Since $R$ is, here, an initial position, it can play the role of \emph{label} for the particle, like a Lagrangian coordinate. If, from the point of view of a static observer at infinity, the particle has an energy $\gamma(R)$, then its free-fall is characterized by the four-velocity
\begin{equation}\label{eq:free-fall_velocity}
u = \frac{\gamma(R)}{A(r)} \partial_t + \sqrt{\gamma^2(R) - A(r)} \, \partial_r.
\end{equation}
One can indeed check that $u$ satisfies the geodesic equation. Let $T$ be the proper time of the particle; integrating $u^r=\dd r/\dd T=\sqrt{\gamma^2(R)-A(r)}$ from the position of the particle at $t=0$ to any moment, we get
\begin{equation}\label{eq:relation_T_R_r}
T-T_0(R) = \int_R^r \frac{\dd \bar{r}}{\sqrt{ \gamma^2(R) - A(\bar{r}) }} ,
\end{equation}
where $T_0(R)$ is the a priori arbitrary origin of $T$ for the particle starting its motion at $R$. This function can be chosen so as to ensure that $u$ is orthogonal to the $T=\mathrm{cst}$ hyper-surfaces,
\begin{equation}
\ddf{T_0}{R} = -\frac{1}{A(R)}\,\sqrt{\gamma^2(R)-A(R)} .
\end{equation}
In the aforementioned conditions, we have $\dd T=u_\mu \dd x^\mu$.

Now consider an infinity of such free-falling particles, filling space, and rewrite the Kottler metric~\eqref{eq:Kottler_metric} using the coordinates $(T,R)$ instead of $(t,r)$. It is not necessary to explicitly integrate Eq.~\eqref{eq:relation_T_R_r}; simply combining $\dd T=-u_\mu \dd x^\mu$ with the expression of $u$, we find
\begin{equation}
\dd s^2 = - \dd T^2 + \frac{1}{\gamma^2(R)} 
\pa{ \left.\pd{r}{R}\right|_T }^2 \dd R^2 
+ r^2(T,R) \, \dd\Omega^2 .
\end{equation}
Furthermore, taking the derivative of Eq.~\eqref{eq:relation_T_R_r} with respect to $R$ (at fixed $T$) leads to
\begin{equation}\label{eq:derivative_r_R}
\left.\pd{r}{R}\right|_T 
= \sqrt{\gamma^2(R)-A(r)}
	\pac{ \frac{\gamma^2(R)}{A(R)\sqrt{ \gamma^2(R)-A(R)}} 
		+ \frac12 \ddf{\gamma^2}{R} \int_R^r \frac{\dd\bar{r}}
							{\pac{\gamma^2(R)-A(\bar{r})}^{3/2}}
			}.
\end{equation}
This generic ``free-fall form'' of the Kottler metric depends on the arbitrary function~$\gamma(R)$, to which one could add the freedom to re-parametrise the $R$ coordinate. Note that the above calculations implicitly assume $\gamma(R)\geq 1$, in other words, all the particles have initially a velocity greater than the escape velocity, so that their Droste radial coordinate $r$ goes from $R$ to infinity. Nevertheless, the same construction is also possible for $\gamma(R)\leq 1$, provided one considers two successive phases of the particles' motion: outgoing first and then ingoing (see Novikov coordinates, at page 826 of Ref.~\cite{1973grav.book.....M}).

Various coordinate systems proposed in the literature, can be recovered from the above construction, by specifying a particular function $\gamma(R)$ and possibly re-parametrising $R$:
\begin{itemize}
\item Lema\^itre coordinates~\cite{1933ASSB...53...51L,1993agns.book..353E} with $\gamma(R)=1$, and $R\mapsto R_{\rm L}$ with the relation $\dd R/\dd R_{\rm L}=A(R)\sqrt{1-A(R)}$. These coordinates were originally considered in the case $\Lambda=0$, which yields $r(T,R_{\rm L})=r_{\rm S}[3 (R_{\rm L}-T)/(2r_{\rm S})]^{2/3}$, and $(\partial r/\partial R_{\rm L})^2 = r_{\rm S}/r$.
\item Robertson coordinates~\cite{1968reco.book.....R} with $\gamma(R)=1$ as well;
\item Novikov coordinates~\cite{1973grav.book.....M} with $\gamma^2(R)=A(R)$, and $R\mapsto R^*$ with $R^*=\sqrt{A(R)/[1-A(R)]}$. Note that one cannot use Eq.~\eqref{eq:derivative_r_R} in this case.
\end{itemize}

Here we keep $\gamma$ fully general, but we introduce $\KK(R)\equiv [1-\gamma(R)]/R^2$, and an inhomogeneous scale factor $\aK(T,R) \equiv r(T,R)/R$, in terms of which the Kottler metric reads
\begin{equation}
\dd s^2 = -\dd T^2 + \aK^2(T,R) \pac{ 
																\pa{1+ \frac{\partial \ln \aK}{\partial \ln R}}^2 \frac{\dd R^2}{1-\KK(R) R^2}
																+ R^2 \, \dd \Omega^2
															} ,
\end{equation}
while scale factor $\aK(T,R)$ satisfies a Friedmann-like equation
\begin{equation}\label{eq:Friedmann_like_equation}
\HK^2
\equiv \pa{ \frac{1}{\aK} \pd{\aK}{T} }^2
= \frac{8\pi G \rhoK_0(R)}{3}\pa{\frac{a_0}{\aK}}^3 - \frac{\KK(R)}{\aK^2} + \frac{\Lambda}{3},
\end{equation}
where $\rhoK_0(R) \define M/[4\pi (a_0 R)^3/3]$ is the mean density of the sphere of radius $a_0 R$. We conclude that each hyper-surface $R=\mathrm{cst}$ behaves exactly as a (layer of a) FL universe, with comoving density $\rhoK_0(R)$ and spatial curvature parameter $\KK(R)$.

%%%%%%%%%%%%%%%%%%%%%%%%%%%%%%%%%%%%%%%%%%%%%%%%%%%%%%%%%%%%%%%%%%%%%
\subsection{Matching the Friedmann-Lema\^itre and Kottler geometries}
%%%%%%%%%%%%%%%%%%%%%%%%%%%%%%%%%%%%%%%%%%%%%%%%%%%%%%%%%%%%%%%%%%%%%

Free-fall coordinates provide a natural extension of cosmic time and comoving coordinates inside the Kottler holes of a SC universe. They also allow us to understand more intuitively the junction between the FL and Kottler space-times at the boundary of a hole. Indeed, as we have seen above, each layer $R=\mathrm{cst}$ expands as a FL universe with density~$\rhoK_0(R)$ and curvature parameter~$\KK(R)$. Hence, if we choose the boundary of a Kottler hole as a sphere of radius $R_{\rm h}$, so that
\begin{equation}\label{eq:junction_condition}
\rho_0 = \rhoK_0(R_{\rm h}) \define \frac{3 M}{4\pi (a_0 R_{\rm h})^3} ,
\end{equation}
and, additionally, set $\gamma$ so that $\KK(R_{\rm h})=K$, then such a sphere will have the same expansion dynamics as the one of the FL cheese. In other words,
\begin{equation}
\forall T \qquad \aK(T,R_{\rm h}) = a(T),
\end{equation}
which matches the Kottler and FL geometries on the layer $R=R_{\rm h}$.

For the sake of completeness, let us also check that, under the conditions specified above, the two Darmois-Israel junction conditions are automatically satisfied. First, the intrinsic metric of the junction hyper-surface (i.e., the hole boundary) is the same whether one computes it from the inside of from the outside,
\begin{align}
\dd s^2_{\rm in}(R=R_{\rm h})
&=
-\dd T^2 + \aK^2(T,R_{\rm h}) R_{\rm h}^2 \dd\Omega^2 \\
&=
-\dd T^2 + a^2(T) R_{\rm h}^2 \dd\Omega^2 \\
&=
\dd s^2_{\rm out}(R=R_{\rm h}).
\end{align}
Secondly, the extrinsic curvature of the junction hyper-surface $R=R_{\rm h}$ is identical whether one computes it from the inside or from the outside. Recall that the extrinsic curvature tensor of a hyper-surface is
\begin{equation}
\mathcal{K}_{ab} \define e\indices{_a^\mu} e\indices{_b^\nu} \nabla_\mu n_\nu,
\end{equation}
where $n$ is a normal unit vector, and the $e_a$ are three tangent vectors to the hyper-surface. Here, the latter can be trivially chosen as $(e_a)=(\partial_T,\partial_\theta,\partial_\ph)$. From the FL (outside) point of view, the unit normal vector reads $n_\mu=a \delta^R_\mu/\sqrt{1-K R_{\rm h}^2}$, from which one deduces
\begin{equation}\label{eq:extrinsic_curvature_FL}
\mathcal{K}_{ab}^{\rm out} \dd x^a \dd x^b
= a(T) R_{\rm h} \sqrt{1 - K R_{\rm h}^2} \, \dd\Omega^2 .
\end{equation}
From the Kottler (inside) point of view, normal vector reads
$n_\mu=\aK\HK R_{\rm h}\delta^R_\mu/\gamma$, from which one computes
\begin{equation}\label{eq:extrinsic_curvature_Kottler}
\mathcal{K}_{ab}^{\rm in} \dd x^a \dd x^b
= \aK(T,R_{\rm h}) R_{\rm h} \sqrt{1 - \KK(R_{\rm h}) R_{\rm h}^2} \, \dd\Omega^2 .
\end{equation}
Thus, both tensors \eqref{eq:extrinsic_curvature_FL} and \eqref{eq:extrinsic_curvature_Kottler} coincide, provided that $\KK(R_{\rm h})=K$ and $\aK(T,R_{\rm h})=a(T)$.

%%%%%%%%%%%%%%%%%%%%%%%%%%%%%%%%
\subsection{Orders of magnitude}
\label{subsec:OOM}
%%%%%%%%%%%%%%%%%%%%%%%%%%%%%%%%

For a SC model to fit with the general philosophy of the DR approximation, it must aim at representing the clumpy, small-scale structure of the Universe. In principle, to be consistent with the typical cross-sectional scale of a light beam associated with astronomical observations, the holes should represent the local environment of individual stars. However, as already discussed in Ref.~\cite{2013PhRvD..87l3526F}, we will not consider such an extreme resolution, but rather stop at the scale of individual galaxies. This leads us to choose the mass parameter of the Kottler regions as $M \sim M_{\rm gal} \sim 10^{11}\,M_\odot$, which corresponds, because of the junction condition~\eqref{eq:junction_condition}, to a typical hole radius
\begin{equation}
R_{\rm h} \sim 1\,\mathrm{Mpc}.
\end{equation}

\begin{figure}[!h]
\centering
\includegraphics[scale=1.1]{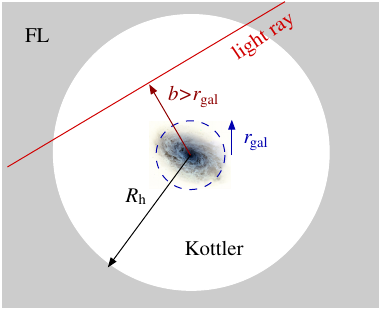}
\caption{Hierarchy of length scale and opacity radius in a Kottler hole.}
\label{fig:OOM}
\end{figure}

A crucial assumption, for the above choice to be meaningful and the calculations of this article to be justified, is that \emph{the clumps at the center of the holes are considered effectively opaque}. In other terms, when studying light propagation through a Swiss cheese in Sec.~\ref{sec:optics_SC} below, we will impose a lower cut-off, for the photon's impact parameter in the Kottler regions, corresponding to the physical size of the central galaxy (see Fig.~\ref{fig:OOM})
\begin{equation}
b > r_{\rm gal} \sim 10\,{\rm kpc} .
\end{equation}
Albeit an intrinsic limitation to the SC approach, such an assumption can be justified statistically (the cross-section of a galaxy is relatively small) and observationally (a galaxy is bright enough to hide a supernova behind it). See Refs.~\cite{2013PhRvD..87l3526F,2009PThPh.122..511O} for further discussions.

Summarizing, Kottler holes are characterized by a hierarchy of length scales
\begin{equation}
r_{\rm S} \ll r_{\rm gal} \ll R_{\rm h} \ll H_0^{-1} 
\lesssim K^{-1/2}, \Lambda^{-1/2},
\end{equation}
essentially controlled by the single dimensionless parameter~$\eps$, so that
\begin{equation}
\eps \define \frac{r_{\rm S}}{R_{\rm h}} 
\sim (H_0 R_{\rm h})^2
\sim 10^{-8}.
\end{equation}
The second relation is deduced from the junction condition \eqref{eq:junction_condition} and the Friedmann equation~\eqref{eq:Friedmann}. Parameter~$\eps$ will be ubiquitous in the perturbative expansions of Sec.~\ref{sec:optics_SC}.

%%%%%%%%%%%%%%%%%%%%%%%%%%%%%%%%%%%%%%%%%%%%%%%%%%%%%%%%%%%%%%%%
\subsection{Backreaction and Swiss-cheese models}
\label{subsec:backreaction_SC}
%%%%%%%%%%%%%%%%%%%%%%%%%%%%%%%%%%%%%%%%%%%%%%%%%%%%%%%%%%%%%%%%

By construction, the Einstein-Straus method allows one to introduce inhomogeneities in a FL universe \emph{without changing its expansion law}. This implies, in particular, that the physical distance between the point masses at the center of two neighbouring holes increases according the Hubble law. In this sense, SC models can be considered backreaction free. In principle, this reasoning should also remain valid for other classes of SC models, a notable representative of which is the LTB SC model, whose holes are filled with an inhomogeneous, non-static, and spherically symmetric dust fluid. While such models differ from the Einstein-Straus one by the choice of the interior metric, the general philosophy is still the same: pick a comoving ball within a FL universe, and reorganize the matter inside it. Again, by construction, this does not change the exterior expansion law (i.e., the expansion law of the FL regions of the SC).

Nevertheless, it might be naive to directly conclude that SC models are backreaction free. Indeed, the \emph{spatially averaged} expansion rate of, e.g., a LTB SC model, can differ from the exterior one~\cite{2011CQGra..28w5002S,2011CQGra..28p4007B,
2013JCAP...12..051L}. Thus, in this sense, SC models are in general \emph{not} backreaction free. We emphasize that this interpretation tacitly considers the averaged expansion rate as the relevant physical quantity to describe the dynamics of the Universe, which is a highly non-trivial, and widely debated assumption. Notable contributions to this debate~\cite{2009JCAP...02..011R,2010JCAP...03..018R,
2012PhRvD..85h3528R,2013JCAP...12..051L}
concluded that, in a fluid-filled and shell-crossing-free universe, the spatially averaged expansion rate really governs the angular distance-redshift relation, and therefore has a powerful physical and observational meaning. However, there is a priori not reason why this result should hold for a more realistic description of the Universe, with shell crossings, formation of virialised structures decoupled from the expansion, etc.

In particular, the class of Swiss-cheese models that we study in the present article, where holes are composed of vacuum and structures in equilibrium, seems precisely to be a counterexample. Indeed, in vacuum there is no unique and natural way to define a 3+1 foliation, and a fortiori a spatially averaged expansion rate. Hence such a notion automatically loses its relevance and its observational meaning---in particular, it cannot drive the distance-redshift relation---when Kottler holes are present. The very meaning of backreaction also becomes unclear, since it usually refers to the influence of inhomogeneities on the average expansion rate. Here, we choose to avoid this issue and identify the expansion rate of our SC model to the one of its FL regions, because it is the only unambiguous choice that we can make. Thus, \emph{from this naive point of view} and according to the discussion of the first paragraph above, the model is backreaction free.

%%%%%%%%%%%%%%%%%%%%%%%%%%%%%%%%%%%%%%%%%%%%%%%%%%%%%%%%%%%%%%%%
%%%%%%%%%%%%%%%%%%%%%%%%%%%%%%%%%%%%%%%%%%%%%%%%%%%%%%%%%%%%%%%%
\section{Geometric optics in Swiss-cheese models}
\label{sec:optics_SC}
%%%%%%%%%%%%%%%%%%%%%%%%%%%%%%%%%%%%%%%%%%%%%%%%%%%%%%%%%%%%%%%%
%%%%%%%%%%%%%%%%%%%%%%%%%%%%%%%%%%%%%%%%%%%%%%%%%%%%%%%%%%%%%%%%

Swiss-cheese models have been used since the late 60s~\cite{1967ApJ...150..737G,1969ApJ...155...89K,
1974ApJ...189..167D,2013PhRvD..87l3526F} to investigate the impact of a clumpy distribution of matter on light propagation, and its consequences on cosmological observables. More recently, they were revisited by replacing Kottler holes by LTB holes, in order to model the large-scale structure of the Universe (voids and walls) rather than its small-scale clumpiness. See Refs.~\cite{2007JCAP...02..013B,2007PhRvD..76l3004M,
2008JCAP...04..008B,2008JCAP...06..021B,2008PhRvD..78h3511V,
2009JCAP...06..010V,2009MNRAS.400.2185C,2011PhRvD..84d4011S,
2012PhRvD..85b3510F} for detailed studies about their optical properties.

In this section, we prove analytically that the DR approximation captures the essential physics of light propagation in SC models with Kottler holes, provided the conditions described in Sec.~\ref{subsec:OOM} are fulfilled.

%%%%%%%%%%%%%%%%%%%%%%%%%%%%%%%%%%%%%%%%%%%%%%%%%%%%%%%%%%%%%%%%
\subsection{Relation between affine parameter and redshift}
\label{subsec:v(z)}
%%%%%%%%%%%%%%%%%%%%%%%%%%%%%%%%%%%%%%%%%%%%%%%%%%%%%%%%%%%%%%%%

The presence of Kottler holes, in a SC universe, modifies the $z(v)$ relation. In this subsection, we show that such a correction is of order $N\eps$, where $N$ is the number of holes crossed by the light beam, and $\eps$ the small parameter defined in Sec.~\ref{subsec:OOM}.

Let us start by investigating the effect of a single hole. Consider a source and an observer comoving with the boundary of the hole (both have a four-velocity $u=\partial_T$); denote respectively ``in'' and ``out'' the emission and the reception events. The redshift of a photon which has travelled through the hole is
\begin{equation}
(1+z)_{\rm in \rightarrow out}
\define
\frac{\nu_{\rm in}}{\nu_{\rm out}}
=
\frac{(u_\mu k^\mu)_{\rm in}}{(u_\mu k^\mu)_{\rm out}}
=
\frac{k_{\rm in}^T}{k_{\rm out}^T}.
\end{equation}
Without any loss of generality, we assume that the photon travels in the plane $\theta=\pi/2$. The symmetries (Killing vectors) of the Kottler geometry imply the existence of two conserved quantities: the ``energy'' $E$ and the ``orbital momentum'' $L$ of the photon, so that, in terms of Droste coordinates,
\begin{equation}
A(r) k^t = E,
\qquad
r^2 k^\ph = L.
\end{equation}
Besides, the coordinate transformation $(t,r) \mapsto (T,R)$ implies
\begin{equation}
k^T 
= 
\gamma \, k^t - \frac{\sqrt{\gamma^2-A}}{A} \, k^r \\
=
\pac{
\gamma \pm \sqrt{\gamma^2-A}\sqrt{1-A\pa{\frac{b}{r}}^2}
}
\frac{E}{A},\label{eq:ktau}
\end{equation}
where $b \define L/E$ is the impact parameter, and $\pm \define \mathrm{sign}(k^r)$ depends on whether the photon is approaching ($-$) or receding ($+$) from the center of the hole. In Eq.~\eqref{eq:ktau}, we have used the constants of motion, and the fact that $k$ is null-like. The redshift is therefore
\begin{equation}\label{eq:redshift_one_hole}
(1+z)_{\rm in \rightarrow out}
=
\frac{A_{\rm out}}{A_{\rm in}}
\frac{
 1 + \sqrt{ 1 - A_{\rm in}/\gamma^2 }
	\sqrt{ 1 - A_{\rm in} \pa{ b/r_{\rm in} }^2 } 
}
{ 1 - \sqrt{ 1-A_{\rm out}/\gamma^2 }
	\sqrt{ 1 - A_{\rm out} \pa{ b/r_{\rm out} }^2 } 
},
\end{equation}
where $A_{\rm in} \define A(r_{\rm in}) = A(a_{\rm in} R_{\rm h})$, and the same for $A_{\rm out}$. This relation is exact. Using the equations which rule the dynamics of the photon and of the hole boundary, it is possible to show that the right-hand side of Eq.~\eqref{eq:redshift_one_hole} is essentially the cosmological redshift $a_{\rm out}/a_{\rm in}$, modulo corrections of order $\eps$ (see Appendix~\ref{app:hole_redshift} for a proof),
\begin{equation}
(1+z)_{\rm in \rightarrow out} 
=
\frac{a_{\rm out}}{a_{\rm in}} \pac{ 1 + \mathcal{O}(\eps) }.
\end{equation}
The corrections hidden in the $\mathcal{O}(\eps)$ term contain both the effect of light deflection in the Kottler hole, and the integrated Sachs-Wolfe (or Rees-Sciama) effect.

If, during its travel through the SC, the photon crosses $N$ holes, then the total redshift is
\begin{equation}\label{eq:redshift_SC}
(1+z)_{\rm s \rightarrow o}
=
\frac{a_{\rm o}}{a_{\rm s}}
\prod_{i=1}^{N} \pac{ 1 + \mathcal{O}(\eps) }
=
\frac{a(T_{\rm o})}{a(T_{\rm s})}
\pac{1 + \mathcal{O}(N \eps)}.
\end{equation}
Equation \eqref{eq:redshift_SC} indicates that if a photon is emitted at cosmic time~$T_{\rm s}$ and observed at $T_{\rm o}$, then the redshift~$z_{\rm SC}$ measured in a SC universe is $z_{\rm FL} + \mathcal{O}(N\eps)$, where $z_{\rm FL}$ is the redshift that would be measured in a FL universe. Interestingly, this \emph{also} implies that the corresponding affine parameters read $v_{\rm SC}=[1 + \mathcal{O}(N\eps)]v_{\rm FL}$. Let us justify this subtle point. By definition, the $T(v)$ relation is governed by
\begin{equation}
\ddf{T}{v} = k^T = \omega = \omega_{\rm o} (1+z),
\end{equation}
thus, because of Eq.~\eqref{eq:redshift_SC},
\begin{equation}
\ddf{v_{\rm SC}}{T} = [1+\mathcal{O}(N\eps)] \ddf{v_{\rm FL}}{T}
\qquad
\text{whence}
\qquad
v_{\rm SC} = [1+\mathcal{O}(N\eps)] v_{\rm FL} .
\end{equation}
We conclude that the affine parameter-redshift relation of a SC only differs by terms of order $N\eps$ from the FL one. This corresponds to the hypothesis \textbf{DR1} of the DR approximation. A numerical illustration, performed by ray tracing in a SC model, is proposed in Sec.~\ref{sec:numerical_results}.

We emphasize that, in the above proof, both the source and the observer were assumed to be \emph{comoving within FL regions}. Hence, two effects which affect the $z(v)$ relation were neglected. First, a source and an observer lying inside Kottler holes would in general undergo a different gravitational potential, depending on their distance to the hole center. The actual redshift must therefore be corrected by a factor $A(r_{\rm o})/A(r_{\rm s})$, which is at most $\sim 1+r_{\rm S}/r_{\rm gal} = 1+\mathcal{O}(100 \eps)$. This effect is therefore sub-dominant when many holes are crossed ($N>100$). The second neglected effect is the one of peculiar velocities (Doppler shift), and is potentially much more significant. Note that it would not only affect the redshift, but also the angular/luminosity distance~\cite{2006PhRvD..73l3526H,
2013PhRvL.110b1302B,2013JCAP...06..002B,2014arXiv1401.3694B}.

For a more general point of view, as already mentioned in Sec.~\ref{subsec:relevance_DR}, we suspect that the deep underlying reason why, here, there is no strong modification of the $v(z)$ relation, is the \emph{absence of backreaction} in SC models. Proving this intuition may however require a dedicated study, whose starting point can be elements proposed in Refs.~\cite{2009JCAP...02..011R,2012MNRAS.426.1121C}.

%%%%%%%%%%%%%%%%%%%%%%%%%%%%%%%%%%%%%%%%%%%%%%%%%%%%%%%%%%%%%%%%
\subsection{Ricci and Weyl focussing in holes}
%%%%%%%%%%%%%%%%%%%%%%%%%%%%%%%%%%%%%%%%%%%%%%%%%%%%%%%%%%%%%%%%

The focussing properties of a Kottler hole are ruled by its optical tidal matrix~$\optical_{\rm K}$. In order to compute it, we first need to specify a Sachs basis. The reference observers' family is chosen as ``generalized comoving observers'', that is, observers with constant Lema\^itre radial coordinate~$R$. As already seen in Sec.~\ref{sec:Swiss-cheese_models}, such observers have the four-velocity $u=\partial_T$ defined by Eq.~\eqref{eq:free-fall_velocity}. The screen vectors $s_1$, $s_2$ form an orthonormal basis of the plane orthogonal to both $u$ and $k$. As before, we can, without loss of generality, assume that the light's trajectory occurs in the equatorial plane $\theta=\pi/2$, so that a first screen vector can be trivially chosen as
\begin{equation}\label{eq:Sachs_basis_Kottler_s1}
s_1 \define \partial_z = -\frac{1}{r} \, \partial_\theta .
\end{equation}
It is straightforward to check that $s_1$ fulfils the transport condition~\eqref{eq:transport_Sachs}. The second screen vector, $s_2$, can then be obtained from the orthogonality and normalization constraints defining the Sachs basis, but it turns out that its explicit expression is not required here.

We now compute the optical tidal matrix~$\optical_{\rm K}$. It is convenient, here, to use the Ricci-Weyl decomposition~\eqref{eq:Ricci-Weyl}. Indeed, since the Kottler geometry describes vacuum, the only contribution to its Ricci tensor is the cosmological constant, $R_{\mu\nu}\propto \Lambda g_{\mu\nu}$, so that
\begin{equation}
\Phi_{00} \define - \frac{1}{2} R_{\mu\nu} k^\mu k^\nu
= 0
\end{equation}
Thus, there is no source of convergence in a Kottler hole, and $\optical_{\rm K}$ is trace free. The calculation of the source of shear~$\Psi_0$ is detailed in Appendix~\ref{app:Weyl_focussing_Kottler}, and the result leads to
\begin{equation}
\optical_{\rm K} =
\begin{pmatrix}
-\Psi_0 & 0 \\
0 & \Psi_0
\end{pmatrix} ,
\qquad \text{with} \qquad
\Psi_0 = \frac{3}{2} \left( \frac{L}{r_{\rm S}^2} \right)^2 \left( \frac{r_{\rm S}}{r} \right)^5.
\end{equation}
As one could expect, the effect of the central mass is to vertically squeeze and horizontally stretch the light beam via tidal forces. The effect is stronger as $M$ increases, and as $b$ decreases. Besides, it is remarkable that the cosmological constant~$\Lambda$, though having an impact on light \emph{deflection}, does not \emph{focus} light. From an observational point of view, it means that for a given value of the affine parameter~$v$, the position on the sky of a light source can be affected by $\Lambda$, but not its magnitude.

The Sachs equation $\ddot{\separation}=\optical_{\rm K}\separation$ can be solved perturbatively~\cite{2013PhRvD..87l3526F} in order to get the expression of the Wronski matrix~$\wronski_{\rm K}$. However, at the order of interest for the discussion of this article, the result is simply
\begin{equation}\label{eq:wronski_Kottler}
\wronski_{\rm K}(\text{out} \leftarrow \text{in})
=
\begin{pmatrix}
\identity_2 & (v_{\rm out} - v_{\rm in}) \identity_2 \\
\zero_2 & \identity_2
\end{pmatrix}
+
\mathcal{O}(\eps).
\end{equation}
In other words, light behaves in the Kottler geometry as in Minkowski space-time, modulo small tidal terms contained in the $\mathcal{O}(\eps)$ term, that we neglect here. Note that neglecting tidal effects, i.e., the source of shear, in the Kottler holes, corresponds to hypothesis~\textbf{DR2} of the DR approximation.

%%%%%%%%%%%%%%%%%%%%%%%%%%%%%%%%%%%%%%%%%%%%%%%%%%%%%%%
\subsection{Effective Ricci focussing in a Swiss cheese}
\label{subsec:effective_Ricci_focussing_SC}
%%%%%%%%%%%%%%%%%%%%%%%%%%%%%%%%%%%%%%%%%%%%%%%%%%%%%%%

\begin{figure}[!h]
\centering
\includegraphics[scale=1.1]{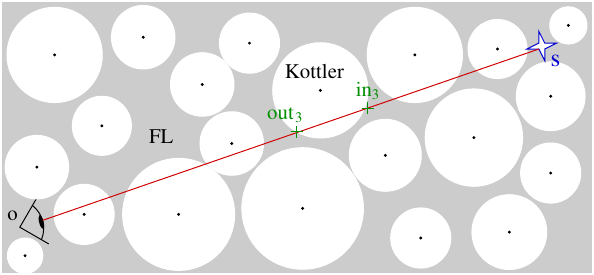}
\caption{A light beam travels, through the Swiss-cheese universe, from a source s to an observer o. The entrance and exit event for the $n$th Kottler hole are respectively denoted $\text{in}_n$ and $\text{out}_n$.}
\label{fig:light_path}
\end{figure}

As already mentioned in Sec.~\ref{subsec:wronski_matrix}, the Wronski matrix is a particularly convenient tool for dealing with a patchwork of space-times, such as a SC model, thanks to its ``Chasles relation''~\eqref{eq:Chasles_Wronski}. Indeed, consider a light beam which travels, in a SC universe, from a source,~s, to an observer,~o, both located in FL regions. If this beam crosses $N$ Kottler holes, then the Wronski matrix describing its evolution can be decomposed as
\begin{equation}
\wronski_{\rm SC}(\text{o}\leftarrow\text{s})
=
\wronski_{\rm FL}(\text{o}\leftarrow\text{out}_N)
\ldots
\underbrace{
\wronski_{\rm FL}(\text{in}_{n+1}\leftarrow\text{out}_n)
\wronski_{\rm K}(\text{out}_n\leftarrow\text{in}_n)
}_{ \wronski_{\rm SC}(\text{in}_{n+1}\leftarrow\text{in}_n)
	\define \wronski_{\rm SC}^{(n)} }
\ldots
\wronski_{\rm FL}(\text{in}_1\leftarrow\text{s})
\end{equation}
where $\text{in}_n$, $\text{out}_n$ respectively denote the entrance and exit of the $n$th hole (see Fig.~\ref{fig:light_path}).

The matrices $\wronski_n \define \wronski_{\rm SC}(\text{in}_{n+1}\leftarrow\text{in}_n)$ represent the elementary bricks of the complete evolution. As we will see below, they merge the FL and Kottler optical properties into an effective behaviour, which coincides with the one proposed by the Dyer-Roeder approximation. First consider the FL part. Since the path between the holes $n$ and $n+1$ is small compared to the cosmological scale $H^{-1}$, one can expand the exact results \eqref{eq:scale_FL}, \eqref{eq:scaledot_FL}, \eqref{eq:jacobi_FL}, and \eqref{eq:jacobidot_FL} to obtain
\begin{equation}\label{eq:wronski_FL}
\wronski_{\rm FL}(\text{in}_{n+1} \leftarrow \text{out}_n)
=
\begin{pmatrix}
\identity_2 & \pac{v_{\rm in}^{(n+1)}-v_{\rm out}^{(n)}} \identity_2 \\
-4\pi G \rho_n \omega_n^2 \pac{v_{\rm in}^{(n+1)}-v_{\rm out}^{(n)}} \identity_2 & \identity_2
\end{pmatrix}
+
\mathcal{O}\pac{\pa{H \Delta T}^2},
\end{equation}
where $\Delta T \define T_{\rm in}^{(n+1)}-T_{\rm out}^{(n)}$.
The matrix product between Eqs.~\eqref{eq:wronski_Kottler} and \eqref{eq:wronski_FL} then yields
\begin{align}
\wronski_{\rm SC}^{(n)}
&=
\identity_4
+
\begin{pmatrix}
\zero_2 & \pac{ v_{\rm in}^{(n+1)} - v_{\rm in}^{(n)} } \identity_2 \\
-4 \pi G \rho_n \omega_n^2 \pac{ v_{\rm in}^{(n+1)} - v_{\rm out}^{(n)} } \identity_2 & \zero_2
\end{pmatrix}
+ \mathcal{O}\pac{\eps,\pa{H \Delta T}^2} \\[2mm]
&=
\identity_4
+
\begin{pmatrix}
\zero_2 & \identity_2  \\
\alpha_n \optical_{\rm FL}(v_n) & \zero_2
\end{pmatrix}
\pac{ v_{\rm in}^{(n+1)} - v_{\rm in}^{(n)} }
+ \mathcal{O}\pac{\eps,\pa{H \Delta T}^2},
\label{eq:Taylor_Wronski_SC}
\end{align}
where we have recognized the FL optical tidal matrix $\optical_{\rm FL}$, given in Eq.~\eqref{eq:optical_FL}, while
\begin{equation}
\alpha_n \define \frac{ v_{\rm in}^{(n+1)} - v_{\rm out}^{(n)} }
				{ v_{\rm in}^{(n+1)} - v_{\rm in}^{(n)} } 
\end{equation}
represents the portion of the path $(\text{in}_{n} \rightarrow \text{in}_{n+1})$ that light spent in the FL region. Interpolating the sequence $(\alpha_n)$ allows one to define a function $\alpha(v)$, which, in principle, depends on the path of light through the Swiss cheese. Note that the way we deal with the expansion~\eqref{eq:Taylor_Wronski_SC} is licit; it is indeed reasonable to consider that the separation between successive holes has the same order of magnitude as the radius of a hole, thus $(H\Delta T)^2 \sim (H R_{\rm h})^2 \sim \eps$ (see Sec.~\ref{subsec:OOM}).

We now show that $\alpha \optical_{\rm FL}$ plays the role of an effective optical tidal matrix. First note that Eq.~\eqref{eq:Taylor_Wronski_SC} can be seen as a first-order Taylor expansion of $\wronski_{\rm SC}(v)$, so that, at leading order in the small parameters of the problem,
\begin{equation}\label{eq:effective_derivative_Wronski}
\begin{pmatrix}
\zero_2 & \identity_2  \\
\alpha(v) \optical_{\rm FL}(v) & \zero_2
\end{pmatrix}
=
\lim_{v'\rightarrow v}
\pd{\wronski_{\rm SC}}{v}(v \leftarrow v')
\define
\pd{\wronski_{\rm SC}}{v}(v \leftarrow v),
\end{equation}
Besides, taking the derivative of the ``Chasles relation'' \eqref{eq:Chasles_Wronski} with respect to $v_3$, and evaluating the result for $v_2=v_3$, yields
\begin{align}
\pd{\wronski_{\rm SC}}{v_3}(v_3 \leftarrow v_1) 
&= \pd{\wronski_{\rm SC}}{v_3}(v_3 \leftarrow v_3)
\wronski_{\rm SC} (v_3 \leftarrow v_1) \\
&\overset{\eqref{eq:effective_derivative_Wronski}}{=}
\begin{pmatrix}
\zero_2 & \identity_2  \\
\alpha(v_3) \optical_{\rm FL}(v_3) & \zero_2
\end{pmatrix}
\wronski_{\rm SC} (v_3 \leftarrow v_1) .
\label{eq:effective_wronski_matrix_equation}
\end{align}
Therefore, comparing the above relation with Eq.~\eqref{eq:master_equation_Wronski} shows that $\alpha \optical_{\rm FL}$ is the effective optical tidal matrix $\optical_{\rm SC}(v)$ for the Swiss cheese. In particular, the Jacobi matrix equation inherited from Eq.~\eqref{eq:effective_wronski_matrix_equation} is
\begin{equation}
\ddot{\jacobi}_{\rm SC} = \alpha \optical_{\rm FL} \jacobi_{\rm SC} .
\end{equation}
This is exactly the hypothesis \textbf{DR3} of the Dyer-Roeder approximation. It also provides a precise definition of the smoothness parameter $\alpha$ in the context of SC models, namely, the \emph{fraction of light path} spent in the FL regions.

%%%%%%%%%%%%%%%%%%%%%%%%%%%%%%
%%%%%%%%%%%%%%%%%%%%%%%%%%%%%%
\section{Numerical results}
\label{sec:numerical_results}
%%%%%%%%%%%%%%%%%%%%%%%%%%%%%%
%%%%%%%%%%%%%%%%%%%%%%%%%%%%%%

This last section aims at illustrating the results of the previous one, using numerical ray tracing in a SC universe.

%%%%%%%%%%%%%%%%%%%%%%%%%%%%%%%%%%%%%%%%%%%%%%%%%%%%%%%%%%%
\subsection{Details of the numerical model and ray-tracing technique}
\label{subsec:ray-tracing_technique}
%%%%%%%%%%%%%%%%%%%%%%%%%%%%%%%%%%%%%%%%%%%%%%%%%%%%%%%%%%%

We consider SC models with a \emph{random} distribution of Kottler holes. As mentioned in Sec.~\ref{subsec:OOM}, each hole is supposed to model the local environment of a galaxy, the central mass being the galaxy itself. Since we do not want all galaxies to have the same mass, we use, in our model, the (stellar) mass function proposed in Ref.~\cite{2004MNRAS.355..764P}, to which we add artificially a factor 10 to take dark matter into account. The result is
\begin{equation}
p(M) \dd M = \frac{1}{\mathcal{N}} \pa{ \frac{M}{10 M^*} }^\alpha \exp\pa{-\frac{M}{10 M^*}} \dd M ,
\end{equation}
with $\alpha=-1.16$, $M^*=7.5 \times 10^{10} \, h^{-2} M_\odot$. This expression is considered valid in the interval $M_{\rm min}<M<M_{\rm max}$, with~\cite{2004MNRAS.355..764P} $M_{\rm min}=10^{8.5}M_\odot$, $M_{\rm max}=10^{13}M_\odot$, and set to zero elsewhere. Thus, the normalization factor~$N$ is
\begin{equation}
\mathcal{N} = \int_{M_{\rm min}}^{M_{\rm max}} \pa{ \frac{M}{10 M^*} }^\alpha \exp\pa{-\frac{M}{10 M^*}} \dd M.
\end{equation}

\begin{figure}[!h]
\centering
\includegraphics[scale=1.1]{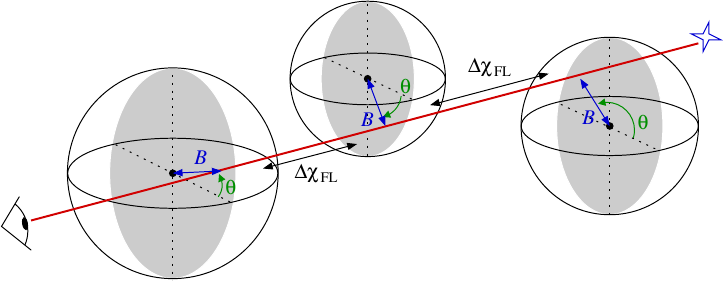}
\caption{Ray tracing in a random SC universe. For each hole, the impact angle~$\theta$, the impact parameter~$B$, and the separation~$\Delta \chi_{\rm FL}$ until the next hole, are random numbers.}
\label{fig:random_SC}
\end{figure}

Regarding ray tracing, the random character of the spatial distribution of holes is modelled using a simple technique where each ray ``creates its own universe''. This method was first proposed by Ref.~\cite{1998PhRvD..58f3501H}, and it has already been used in many studies involving SC models, see, e.g., Refs.~\cite{2008JCAP...04..008B,
2009PhRvD..80l3020K,
2011PhRvD..84d4011S,
2012PhRvD..85b3510F}. It consists in putting holes on the light's trajectory, with random (comoving areal\footnote{The usual impact parameter~$b=L/E$ is defined with respect to the Droste coordinate system. Its comoving counterparts are $\beta=b/a_{\rm in}$ and $B=f_K(\beta)$. Note that, in practice, $B\approx\beta$ since $\sqrt{|K|}\beta \sim b H_0 \ll 1$.\label{footnote:impact_parameter}}) impact parameter~$B$, random impact angle $\theta$, and with a random comoving length~$\Delta\chi_{\rm FL}$ between the exit of the $n$th hole and the entrance of the $(n+1)$th one. The situation is depicted in Fig.~\ref{fig:random_SC}.

We consider that all the impacts positions, within the authorized cross section of a given hole, are equi-probable. Thus, the random impact angle~$\theta$ is uniformly distributed; and the probability density function (PDF) of the impact parameter~$B$ reads
\begin{equation}\label{eq:PDF_B}
p(B)\,\dd B = \frac{2 B\,\dd B}{R_{\rm h}^2 - R_{\rm gal}^2},
\qquad
R_{\rm gal} < B < R_{\rm h},
\end{equation}
where $R_{\rm h}$ is the comoving areal radius of the hole---related to its central mass via Eq.~\eqref{eq:junction_condition}---and $R_{\rm gal}$ is the opacity radius mentioned in Sec.~\ref{subsec:OOM}. We choose to link it to the mass~$M$ of the galaxy via a constant density~$\rho_{\rm gal} = 5 \times 10^6 \, M_\odot \, \mathrm{kpc}^{-3}$, so that
\begin{equation}
R_{\rm gal}(M) \define \pa{ \frac{3 M}{4\pi\rho_{\rm gal}} }^{1/3}
= \pa{ \frac{\rho_0}{\rho_{\rm gal}} }^{1/3} R_{\rm h}(M).
\end{equation}

As a last simplifying assumption, the FL separation~$\Delta\chi_{\rm FL}$ between two successive holes is also chosen to be uniformly distributed\footnote{Note that this \emph{does not} correspond to a Swiss-cheese model with randomly distributed, non-overlapping, holes. Strictly speaking, in the latter situation, there would be a correlation between the impact parameter $B$ and $\Delta\chi_{\rm FL}$, because, e.g., $\Delta\chi_{\rm FL}=0$ is only possible between two holes with the same impact parameter. We do not take this correlation into account for simplicity.} between $0$ and $\max(\Delta\chi_{\rm FL})=2\mean{\Delta\chi_{\rm FL}}$. We parametrize the mean value with an effective constant smoothness parameter~$\bar{\alpha}$, so that
\begin{equation}
\mean{\Delta\chi_{\rm FL}} 
= \frac{\bar{\alpha}}{1-\bar{\alpha}} \, \mean{\Delta\chi_{\rm K}},
\end{equation}
where $\mean{\Delta\chi_{\rm K}}$ is the comoving distance spent inside a Kottler hole. The calculation of this quantity is given in Appendix~\ref{app:mean_Delta_chi_K}, and the result is
\begin{equation}
\mean{\Delta\chi_{\rm K}} 
\approx \frac{4}{3} \pa{ \frac{3}{4\pi\rho_0} }^{1/3} 
\int_{M_{\rm min}}^{M_{\rm max}} 
p(M) M^{1/3} \; \dd M .
\end{equation}

In practice, the author wrote a Mathematica program to perform ray tracing in the conditions described previously. Calculations start at the observation event and go backward in time. The code consists in iterating the following steps. (i) Pick a FL comoving distance $\Delta\chi_{\mathrm{FL},n}$ and propagate the beam across it; (ii) pick a mass~$M_n$, an impact parameter~$B_n$, and an impact angle~$\theta_n$ defining light propagation through the $n$th Kottler hole; (iii) compute the redshift and Wronski matrix across this hole. We stress that, for those numerical calculations, we did not use the lowest-order expression~\eqref{eq:wronski_Kottler} for the Wronski matrix~$\wronski_{\rm K}$, but rather the one of Ref.~\citep{2013PhRvD..87l3526F}, which takes into account tidal effects at order one.

%%%%%%%%%%%%%%%%%%%%%%%%%%%%%%%%%%%%%%%%%%%%%%%%%%%%%%%%%%%
\subsection{Relation between affine parameter and redshift}
%%%%%%%%%%%%%%%%%%%%%%%%%%%%%%%%%%%%%%%%%%%%%%%%%%%%%%%%%%%

\begin{figure}[!h]
\centering
\includegraphics[scale=1.6]{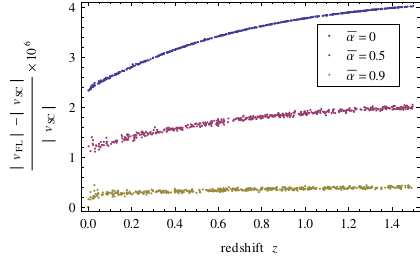}
\caption{Relative difference between the affine parameter-redshift relation $|v_{\rm FL}(z)|$ of a FL model and of a SC model $|v_{\rm SC}(z)|$, with different values for the mean smoothness parameter $\bar{\alpha}$. From top to bottom, $\bar{\alpha}=0$ (blue), $\bar{\alpha}=0.5$ (magenta), and $\bar{\alpha}=0.9$ (yellow). Absolute values are used in order to avoid any conventional discussions about whether $v$ increases or decreases towards the past.}
\label{fig:affine_parameter_redshift}
\end{figure}

In this paragraph, we illustrate the results of Sec.~\ref{subsec:v(z)}, regarding the affine parameter-redshift relation. Figure~\ref{fig:affine_parameter_redshift} shows the relative difference, for the $v(z)$ relation, between a FL model and three different SC models, from very clumpy ($\bar{\alpha}=0$) to very smooth ($\bar{\alpha}=0.9$). All the models are characterized by the cosmological parameters obtained by the Planck experiment~\cite{2013arXiv1303.5076P}, namely $\Omega_{\rm m0}=0.315$, $\Omega_{\Lambda 0}=0.685$. For each SC model, 500 observations are simulated within the range $0<z<1.5$, according to the method presented in Sec.~\ref{subsec:ray-tracing_technique}.

Even for a model entirely filled by Kottler holes ($\bar{\alpha}=0$), we see that the relative correction to the $v(z)$ relation is very small, less than $10^{-5}$. This order of magnitude is compatible with the results of Sec.~\ref{subsec:v(z)}.

%%%%%%%%%%%%%%%%%%%%%%%%%%%%%%%%%%%%%%%%%%%%%%%%%%%
\subsection{Relation between distance and redshift}
%%%%%%%%%%%%%%%%%%%%%%%%%%%%%%%%%%%%%%%%%%%%%%%%%%%

\begin{figure}[!h]
\centering
\includegraphics[scale=1.6]{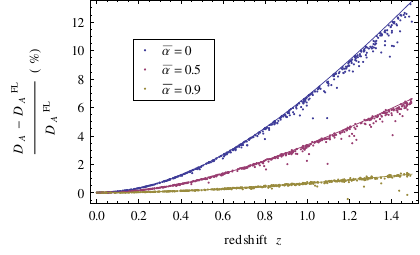}
\caption{Correction, with respect to FL, of the angular distance-redshift relation $D_{\rm A}(z)$ of several models. Dots are simulated observations in SC models with three different mean smoothness parameters $\bar{\alpha}$. From top to bottom, $\bar{\alpha}=0$ (blue), $\bar{\alpha}=0.5$ (magenta), and $\bar{\alpha}=0.9$ (yellow). Solid lines indicate the corresponding DR relations $D_{\rm A}^{\rm DR}(z)$ with constant smoothness parameter $\alpha=\bar{\alpha}$.}
\label{fig:distance_redshift}
\end{figure}

In this paragraph, we illustrate the results of Sec.~\ref{subsec:effective_Ricci_focussing_SC}, regarding effective Ricci focussing in a SC model, and its comparison with the DR approximation. Figure~\ref{fig:distance_redshift} shows the relative correction to the $D_{\rm A}(z)$ relation, for three different SC and DR models, with respect to the corresponding FL models. As before, the cosmological parameters are Planck's best-fit ones, and for each SC model, 500 observations are simulated within the range $0<z<1.5$.
%We stress that those simulations take into account the effect of Weyl focussing inside the Kottler holes. For that purpose, we used the results of Ref.~\cite{2013PhRvD..87l3526F} for $\wronski_{\rm K}$.

First note that the difference between $D_{\rm A}^{\rm SC}$ and $D_{\rm A}^{\rm FL}$ is of the percent order, and reaches more than $12\%$ at $z=1.5$ for a very clumpy SC model (the cosmological implications of this difference are discussed in Refs.~\cite{2013PhRvD..87l3526F,2013PhRvL.111i1302F}). It confirms that, in SC models, the correction to the $v(z)$ relation (Fig.~\ref{fig:affine_parameter_redshift}) is negligible compared to the $D_{\rm A}(v)$ one.

In Fig.~\ref{fig:distance_redshift}, the good agreement between dots and solid lines numerically confirms the main point of this article, namely, that the Dyer-Roeder approximation provides a good effective description of light propagation in SC models. However, this agreement is not perfect, especially for $\bar{\alpha}=0$, where the mean behaviour of $D_{\rm A}^{\rm SC}(z)$ is slightly overestimated by $D_{\rm A}^{\rm DR}(z)$, with some rare events in strong disagreement. As we shall see in the next subsection, this is due to the neglected Weyl lensing effects, i.e., departures from hypothesis \textbf{DR2}. 

%%%%%%%%%%%%%%%%%%%%%%%%%%%%%%%%%%%%%%%%%%%%%%%%%%%%%%%%%%%%%%
\subsection{Lensing beyond the Dyer-Roeder approximation}
%%%%%%%%%%%%%%%%%%%%%%%%%%%%%%%%%%%%%%%%%%%%%%%%%%%%%%%%%%%%%%

This last subsection is dedicated to some lensing effects which are present in a SC model, but not taken into account by the DR approximation. In order to compare the focussing properties of a given space-time with those of FL model, it is convenient to introduce the amplification (or magnification) matrix
\begin{equation}
\amplification
\define
\jacobi \cdot \jacobi_{\rm FL}^{-1}
= \frac{\jacobi}{\omega_{\rm o} D_{\rm A}^{\rm FL}} .
\end{equation}
This matrix describes the geometrical transformations of an image (magnification, deformation, rotation) which add to the global FL focussing effect. For instance, the relative magnification~$\mu$, defined as the ratio between observed angular size of an object, and the one that would be observed in a FL universe, is related to $\amplification$ via\footnote{Note by the way that the usual names ``amplification'' or ``magnification'' matrix for $\amplification$ are particularly misleading, and would be much more adapted to $\amplification^{-1}$.}
\begin{equation}
\mu 
\define
\frac{\dd \Omega_{\rm o}^2}{\dd \Omega_{\rm o, FL}^2}
=
\pa{ \frac{D_{\rm A}^{\rm FL}}{D_{\rm A}} }^2
=
\frac{1}{\det\amplification }.
\end{equation}
In general, as any $2\times 2$ matrix, $\amplification$ can be decomposed as the product between an SO(2) matrix, encoding the image rotation; and a symmetric matrix, encoding its distortion:
\begin{equation}
\amplification
=
\begin{pmatrix}
\cos\psi & \sin\psi \\
-\sin\psi & \cos\psi
\end{pmatrix}
\begin{pmatrix}
1-\kappa-\gamma_1 & -\gamma_2 \\
-\gamma_2 & 1-\kappa+\gamma_1
\end{pmatrix},
\end{equation}
where
\begin{equation}
\psi = \arctan\pa{ 
				\frac{ \mathcal{A}_{12} - \mathcal{A}_{21} }
					{\mathcal{A}_{11} + \mathcal{A}_{22} } 
				}
\end{equation}
is the rotation angle, $\kappa$ is the convergence, and $\gamma=\gamma_1+\mathrm{i}\gamma_2$ the shear, of the image. It is straightforward to check that the magnification is related to those quantities according to
\begin{equation}\label{eq:magnification_convergence_shear}
\mu = \frac{1}{(1-\kappa)^2 - |\gamma|^2}.
\end{equation}

\begin{figure}[!h]
\centering
\includegraphics[width=\columnwidth]{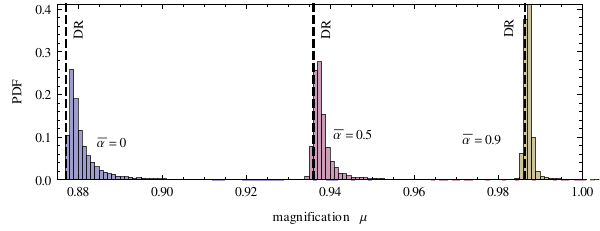}\\
\vspace*{0.5cm}
\includegraphics[width=\columnwidth]{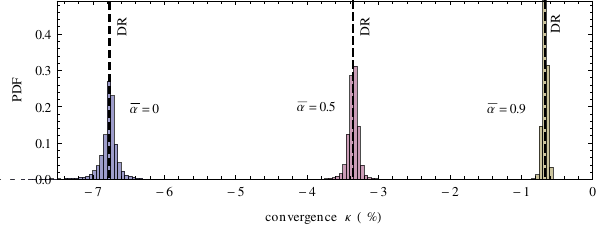}\\
\vspace*{0.5cm}
\includegraphics[width=0.49\columnwidth]{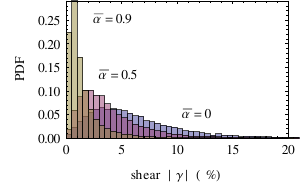}
\includegraphics[width=0.49\columnwidth]{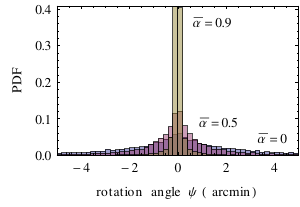}
\caption{Probability density functions (PDFs) of the magnification $\mu$ (top panel), convergence $\kappa$ (middle panel), shear $|\gamma|$ (bottom-left panel), and rotation angle $\psi$ (bottom-right panel), in three different SC models with respective mean smoothness parameter $\bar{\alpha}=0$ (blue), $\bar{\alpha}=0.5$ (magenta), and $\bar{\alpha}=0.9$ (yellow). The magnification and convergence predicted by the DR approximation are also indicated, for comparison, by vertical dashed lines.}
\label{fig:PDFs}
\end{figure}

\begin{figure}[!h]
\centering
\includegraphics[width=0.49\columnwidth]{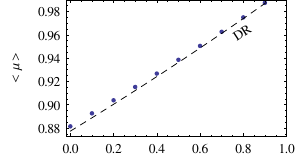}
\includegraphics[width=0.49\columnwidth]{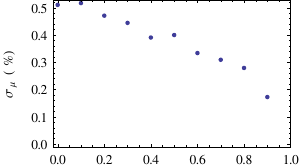}\\
\vspace*{0.5cm}
\includegraphics[width=0.49\columnwidth]{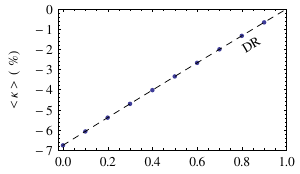}
\includegraphics[width=0.49\columnwidth]{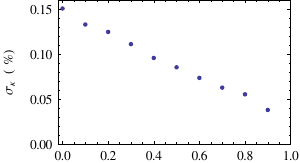}\\
\vspace*{0.5cm}
\includegraphics[width=0.49\columnwidth]{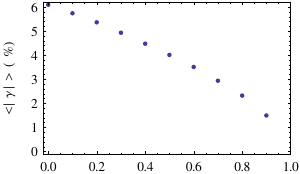}
\includegraphics[width=0.49\columnwidth]{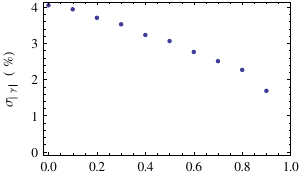}\\
\vspace*{0.5cm}
\includegraphics[width=0.49\columnwidth]{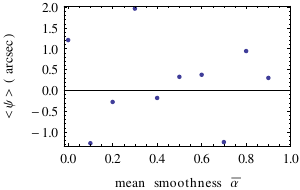}
\includegraphics[width=0.49\columnwidth]{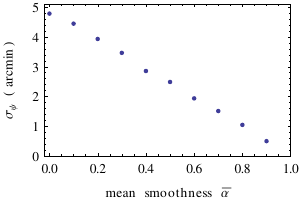}
\caption{First two moments---mean (left column) and standard deviation (right column)---of the PDFs of, from top to bottom, the magnification $\mu$, convergence $\kappa$, shear $|\gamma|$, and rotation angle $\psi$, in SC models, as a function of their mean smoothness parameter $\bar{\alpha}$. The magnifications and convergences obtained from the DR approach are indicated by dashed lines, for comparison.}
\label{fig:moments}
\end{figure}

In the DR approximation, shear and rotation are neglected. But since we are able to compute them numerically for SC models, it is interesting to see how they can induce a departure from the DR behaviour. Figure~\ref{fig:PDFs} shows, as examples, the PDFs of the optical quantities, generated by simulating $10^4$ observations at redshift $z=1$ in three different SC models with $\bar{\alpha}=0$, $0.5$, $0.9$. The values predicted by the DR approximation, with $\alpha=\bar{\alpha}$, are indicated for comparison. The evolution of the first two moments of the PDFs (mean and standard deviation) with the mean smoothness parameter~$\bar{\alpha}$ of the SC model are depicted in Fig.~\ref{fig:moments}. In this figure, the mean magnification~$\langle\mu\rangle$ and convergence~$\langle\kappa\rangle$ are also compared with the DR values.

We see that the DR approximation predicts a value for the convergence in excellent agreement with the mean convergence $\langle\kappa\rangle$ in SC models, but slightly underestimates the mean magnification $\langle\mu\rangle$, as already suspected in Fig.~\ref{fig:distance_redshift}. The difference increases as the mean smoothness parameter $\bar{\alpha}$ decreases, and reaches $\langle\mu\rangle-\mu_{\rm DR}=0.4\%$ for $\bar{\alpha}=0$. More precisely, we see from the top panel of Fig.~\ref{fig:PDFs} that $\mu_{\rm DR}$ gives essentially the \emph{most probable} magnification, which is different from the \emph{mean} magnification because the PDF is clearly skewed. Besides, since the PDF of the convergence seems much more symmetric, such a skewness can only come from the shear. Thus, we conclude that, in SC models, departures from the DR behaviour are due to neglecting Weyl lensing, i.e. hypothesis \textbf{DR2}.

However, such departures remain small, since in the worst case $\langle\mu\rangle-\mu_{\rm DR}=0.4\%$, while $\langle\mu\rangle-\mu_{\rm FL}=\langle\mu\rangle-1=-12\%$. This could be surprising, because \emph{the shear is not intrinsically negligible} compared to the convergence, we indeed see from Fig.~\ref{fig:moments} that $\langle\kappa\rangle\sim\langle|\gamma|\rangle\sim\%$. The difference between those optical quantities is that, fortunately, the magnification $\mu$ involves $\kappa$ at order one, but $\gamma$ only at order two [see Eq.~\eqref{eq:magnification_convergence_shear}]. This justifies a posteriori the expression~\eqref{eq:wronski_Kottler} of $\wronski_{\rm K}$, used in the proof of Sec.~\ref{subsec:effective_Ricci_focussing_SC}, where we completely dropped the Weyl focussing effects. Such an approximation would not have been consistent if we were interested in anything else than the angular distance, i.e. the determinant of the Jacobi matrix.

%%%%%%%%%%%%%%%%%%%%%%%%%%%%%%%%%%%%%%%%%%%%%%%%%%%%%%%%%%%%%%%%
%%%%%%%%%%%%%%%%%%%%%%%%%%%%%%%%%%%%%%%%%%%%%%%%%%%%%%%%%%%%%%%%
\section*{Conclusion}
\addcontentsline{toc}{section}{Conclusion}
%%%%%%%%%%%%%%%%%%%%%%%%%%%%%%%%%%%%%%%%%%%%%%%%%%%%%%%%%%%%%%%%
%%%%%%%%%%%%%%%%%%%%%%%%%%%%%%%%%%%%%%%%%%%%%%%%%%%%%%%%%%%%%%%%

In this article, we analysed the suspected correspondence between light propagation in Einstein-Straus Swiss-cheese (SC) models and the Dyer-Roeder (DR) approximation. Invoking both analytical proofs and numerical illustrations, we proved that such an approximation is indeed excellent for predicting the distance-redshift relation of SC models, provided that (1) the matter clumps at the center of SC holes are effectively opaque, and (2) reasonable orders of magnitude are taken for the mass and compactness of the clumps.

Rather than just checking the good agreement between the results of both approaches, our main purpose was to understand why the various hypotheses of the DR approximation are satisfied in SC models. It appeared that:
\begin{itemize}
\item The affine parameter-redshift relation $v(z)$ is essentially the same in SC and FL models because the deflection and ISW effects are negligible. Independently of such effects, we also suspect that the absence of backreaction in our SC model (the holes do not affect the expansion law of the FL regions) is the deep reason why the FL $v(z)$ relation holds.
\item In SC models, Weyl lensing (source of shear and rotation) and Ricci lensing (source of convergence) are intrinsically comparable. However, compared to the latter, the former have a negligible impact on the angular distance-affine parameter relation $D_{\rm A}(v)$, because shear only appears at order two in the expression of the magnification.
\item The way the DR approximation deals with Ricci lensing, i.e., making heuristically the replacement $\rho \rightarrow \alpha \rho$ in the Sachs equation, works in SC models because (i) the clumps inside the holes are considered opaque; and (ii) FL regions and Kottler regions alternate many times over cosmological scales. This, indeed, allows the SC Wronski matrix to get an effective behavior which fits the DR one.
\end{itemize}

In the case of extremely clumpy SC models (entirely filled by Kottler holes), small departures from the DR predictions are observed, regarding the mean magnification. We saw that they were due to the effect of the neglected Weyl lensing. However, such departures remain small, since at worst $\langle \mu \rangle - \mu_{\rm DR} = 0.4\%$, to be compared with $\langle \mu \rangle - \mu_{\rm FL} = -12\%$. Moreover, the PDF of the magnification in SC models being skewed, the most probable magnification is smaller than the mean one, and thus in even better agreement with $\mu_{\rm DR}$. We conclude that, regarding the distance redshift relation, one can safely consider the DR and SC approaches as \emph{equivalent}.

The question of whether those approaches are relevant alternatives to the standard interpretation cosmological data is beyond the scope of this article. It regroups at least two crucial issues of modern cosmology. The first one is the amplitude of backreaction, neglected in both the DR and SC approaches. The second one concerns the actual clumpiness of our Universe, which is closely related to the problem of structure formation, and even to the question of the nature of dark matter.

%%%%%%%%%%%%%%%%%%%%%%%%%%%%%%%%%%%%%%%%%%%%%%%%%%%%%%%%%%%%%%%%
%%%%%%%%%%%%%%%%%%%%%%%%%%%%%%%%%%%%%%%%%%%%%%%%%%%%%%%%%%%%%%%%
\section*{Acknowledgements}
%%%%%%%%%%%%%%%%%%%%%%%%%%%%%%%%%%%%%%%%%%%%%%%%%%%%%%%%%%%%%%%%
%%%%%%%%%%%%%%%%%%%%%%%%%%%%%%%%%%%%%%%%%%%%%%%%%%%%%%%%%%%%%%%%

The author wishes to warmly thank his scientific family, namely Cyril Pitrou and Jean-Philippe Uzan, for useful discussions and remarks on early versions of this article. Must also be acknowledged, for their comments, Thomas Buchert, Jean Eisenstaedt, Giovanni Marozzi, Fabien Nugier, and Syksy R\"as\"anen. I thank Andreas Finke for pointing out a mistake in Sec.~\ref{subsec:free-fall_Kottler} in the published version. This work made in the ILP LABEX (under reference ANR-10-LABX-63) was supported by French state funds managed by the ANR within the Investissements d'Avenir programme under reference ANR-11-IDEX-0004-02.

%%%%%%%%%%%%%%%%%%%%%%%%%%%%%%%%%%%%%%%%%%%%%%%%%%%%%%%%%%%%%%%%
%%%%%%%%%%%%%%%%%%%%%%%%%%%%%%%%%%%%%%%%%%%%%%%%%%%%%%%%%%%%%%%%
%%%%%%%%%%%%%%%%%%%%%%%%%%%%%%%%%%%%%%%%%%%%%%%%%%%%%%%%%%%%%%%%

\appendix

%%%%%%%%%%%%%%%%%%%%%%%%%%%%%%%%%%%%%%%%%%%%%%%%%%%%%%%%%%%%%%%%
%%%%%%%%%%%%%%%%%%%%%%%%%%%%%%%%%%%%%%%%%%%%%%%%%%%%%%%%%%%%%%%%
\section{Redshift through a Kottler hole}
\label{app:hole_redshift}
%%%%%%%%%%%%%%%%%%%%%%%%%%%%%%%%%%%%%%%%%%%%%%%%%%%%%%%%%%%%%%%%
%%%%%%%%%%%%%%%%%%%%%%%%%%%%%%%%%%%%%%%%%%%%%%%%%%%%%%%%%%%%%%%%

Let us show that the redshift of a photon crossing a Kottler hole is essentially $a_{\rm out}/a_{\rm in}$,
\begin{equation}
(1+z)_{\rm in \rightarrow out}
=
\frac{a_{\rm out}}{a_{\rm in}}
\times
\underbrace{
\frac{A_{\rm out}}{A_{\rm in}}
\frac{
 r_{\rm in} + \sqrt{ 1 - A_{\rm in}/\gamma^2 }
	\sqrt{ r_{\rm in}^2 - b^2 A_{\rm in} } 
	}
{ r_{\rm out} - \sqrt{ 1-A_{\rm out}/\gamma^2 }
	\sqrt{ r_{\rm out}^2 - b^2 A_{\rm out} } 
}
			}_{ 1+\mathcal{O}(\eps) },
\label{eq:z_hole_exact}
\end{equation}
where $\eps \define r_{\rm S}/R_{\rm h}$. To do so, we must use both the dynamics of the photon $r_{\rm p}(t)$ and of the hole boundary $r_{\rm h}(t)$, in terms of Droste coordinates:
\begin{align}
\ddf{r_{\rm p}}{t}
&= \pm A(r_{\rm p}) 
	\sqrt{1-A(r_{\rm p})\pa{\frac{b}{r_{\rm p}}}^2}, 
\label{eq:dynamics_photon}
\\
\ddf{r_{\rm h}}{t}
&= A(r_{\rm h}) 
\sqrt{ 1 - \frac{A(r_{\rm h})}{\gamma^2} },
\label{eq:dynamics_hole}
\end{align}
where the $\pm$ sign depends on whether the photon is approaching ($-$) or receding ($+$) from the hole center. The order of magnitude of the time spent by the photon inside a hole is the radius of the latter, $\Delta t \sim r_{\rm h}$. From Eq.~\eqref{eq:dynamics_hole}, we deduce that, during this amount of time, the hole radius increases by $\delta r_{\rm h}/r_{\rm h} \sim \sqrt{\eps}$. The corresponding variation of $A(r_{\rm h})$ is then $A_{\rm out}/A_{\rm in}-1 \sim \eps^{3/2}$. Hence, since we aim at studying the expression of $(1+z)_{\rm in \rightarrow out}$ up to order one in $\eps$, we can already neglect the ratio $A_{\rm out}/A_{\rm in}$ which appears in Eq.~\eqref{eq:z_hole_exact}.

Let $t_{\rm m}$ be the instant when the coordinate distance between the photon and the center of the hole is minimal, $r_{\rm p}(t_{\rm m})=r_{\rm m}$. Taylor-expanding the function $r_{\rm h}(t)$ from $t_{\rm in}$ to $t_{\rm m}$ leads to
\begin{equation}
r_{\rm h}(t_{\rm m})
= r_{\rm in} 
	+ (t_{\rm m}-t_{\rm in}) 
	A_{\rm in} \sqrt{ 1 - \frac{A_{\rm in} }{\gamma^2} } 
	+ r_{\rm in}\mathcal{O}(\eps),
\label{eq:expansion_rh}
\end{equation}
where we replaced $(\dd r_{\rm h}/\dd t)_{\rm in}$ by its expression~\eqref{eq:dynamics_hole}. Besides, from Eq.~\eqref{eq:dynamics_photon}, we get
\begin{equation}
t_{\rm m}-t_{\rm in}
= 
\int_{r_{\rm m}}^{r_{\rm in}} 
\frac{\dd r}
{A \sqrt{1-(b/r)^2A} }
=
\frac{ \sqrt{r_{\rm in}^2 - b^2 A_{\rm in}} }{ A_{\rm in} }
+
\underbrace{
\int_{r_{\rm m}}^{r_{\rm in}} 
\frac{r^2 - b^2 A/2}{  \sqrt{r^2 - b^2 A} }
\frac{A' \dd r}{2 A^2}
}_{\delta_{\rm in}},
\label{eq:delta_t_in}
\end{equation}
where the second equality is an integration by parts. A rough analysis shows that $\delta_{\rm in}=(r_{\rm in}^3/b^2) \mathcal{O}(\eps)$, that is, using the orders of magnitude of Sec.~\ref{subsec:OOM}, $\delta_{\rm in}=r_{\rm in} \mathcal{O}(\eps^{1/2})$. Hence, we conclude that Eq.~\eqref{eq:expansion_rh} can be rewritten as
\begin{equation}
r_{\rm h}(t_{\rm m}) = 
r_{\rm in} + 
\sqrt{ r_{\rm in}^2 - b^2 A_{\rm in} }
\sqrt{ 1 - \frac{A_{\rm in} }{\gamma^2} }
+ r_{\rm in} \mathcal{O}(\eps).
\end{equation}
The same calculations, but starting from an expansion of $r_{\rm h}(t)$ from $t_{\rm in}$ to $t_{\rm m}$, give
\begin{equation}
r_{\rm h}(t_{\rm m}) = 
r_{\rm out} -
\sqrt{ r_{\rm out}^2 - b^2 A_{\rm out} }
\sqrt{ 1 - \frac{A_{\rm out} }{\gamma^2} }
+ r_{\rm out} \mathcal{O}(\eps),
\end{equation}
so that, finally,
\begin{equation}
\frac{
 r_{\rm in} + \sqrt{ 1 - A_{\rm in}/\gamma^2 }
	\sqrt{ r_{\rm in}^2 - b^2 A_{\rm in} } 
	}
{ r_{\rm out} - \sqrt{ 1-A_{\rm out}/\gamma^2 }
	\sqrt{ r_{\rm out}^2 - b^2 A_{\rm out} } 
}
=
\frac{r_{\rm h}(t_{\rm m}) + r_{\rm in} \mathcal{O}(\eps) }
{r_{\rm h}(t_{\rm m}) + r_{\rm out} \mathcal{O}(\eps)}
=
1+\mathcal{O}(\eps).
\end{equation}
%

%%%%%%%%%%%%%%%%%%%%%%%%%%%%%%%%%%%%%%%%%%%%%%%%%%%%%%%%%%%%%%%%
%%%%%%%%%%%%%%%%%%%%%%%%%%%%%%%%%%%%%%%%%%%%%%%%%%%%%%%%%%%%%%%%
\section{Source of shear in Kottler geometry}
\label{app:Weyl_focussing_Kottler}
%%%%%%%%%%%%%%%%%%%%%%%%%%%%%%%%%%%%%%%%%%%%%%%%%%%%%%%%%%%%%%%%
%%%%%%%%%%%%%%%%%%%%%%%%%%%%%%%%%%%%%%%%%%%%%%%%%%%%%%%%%%%%%%%%

We compute the Weyl part (source of shear) of the optical tidal matrix~$\optical_{\rm K}$ for the Kottler geometry, using the regular Droste coordinates $(t,r,\theta,\ph)$. The non-zero components of the Riemann tensor are
\begin{gather}
R_{trtr} = \frac{A'}{2}, \qquad
R_{t\theta t\theta} = \frac{r A' A}{2}, \qquad
R_{t\ph t\ph} = \sin^2\theta R_{t\theta t\theta}, \\
R_{r\theta r\theta} = -\frac{r A'}{2 A}, \qquad
R_{r\ph r\ph} = \sin^2\theta R_{r\theta r\theta}, \qquad
R_{\theta\ph\theta\ph} = r^2 (1-A) \sin^2\theta.
\end{gather}
Without loss of generality, we assume that the axes have been chosen so that the light path lies in the plane $\theta=\pi/2$, which implies $k^\theta=0$. The four-velocity of the reference observers is given by Eq.~\eqref{eq:free-fall_velocity}, in particular $u^\theta=0$.
\begin{align}
\mathcal{R}^{\rm K}_{11} 
&= R_{\mu\nu\alpha\beta}s_1^\mu k^\nu k^\alpha s_1^\beta \\
&= - r^{-2} R_{\mu\theta\nu\theta} k^\mu k^\nu \\
&= - r^{-2} \pac{ \frac{r A' A}{2} (k^t)^2 
					- \frac{r A'}{2 A} (k^r)^2 
					+ r^2 (1-A) (k^\ph)^2 }.
\end{align}
but
\begin{equation}
\frac{r A' A}{2} (k^t)^2 - \frac{r A'}{2 A} (k^r)^2
= -\frac{r A'}{2} \pac{ g_{tt} (k^t)^2 + g_{rr} (k^r)^2}
= \frac{r A'}{2} g_{\ph\ph} (k^\ph)^2
= \frac{r^3 A'}{2} (k^\ph)^2
\end{equation}
therefore,
\begin{equation}
\mathcal{R}^{\rm K}_{11} 
= \pac{ -\frac{r A'}{2} - (1-A) } (k^\ph)^2
= -\frac{3 r_{\rm S}}{2} \frac{L^2}{r^5} .
\end{equation}
Since the Ricci-focussing term is zero, the optical tidal matrix is trace-free, so that $\mathcal{R}^{\rm K}_{11}=-\mathcal{R}^{\rm K}_{22}$. Besides, the off-diagonal terms $\mathcal{R}^{\rm K}_{12}=\mathcal{R}^{\rm K}_{21}$ are zero, indeed
\begin{equation}
\mathcal{R}^{\rm K}_{12}
\propto R_{\theta\nu\alpha\beta} k^\nu k^\alpha s_2^\beta,
\end{equation}
and the vectors $k$, $s_2$ have no components along $\partial_\theta$ (so that $\nu,\alpha,\beta \not= \theta$), while all the components of the Riemann tensor involving a single index $\theta$ vanish.

%%%%%%%%%%%%%%%%%%%%%%%%%%%%%%%%%%%%%%%%%%%%%%%%%%%%%%%%%%%%%%%%
%%%%%%%%%%%%%%%%%%%%%%%%%%%%%%%%%%%%%%%%%%%%%%%%%%%%%%%%%%%%%%%%
\section{Mean Kottler path}
\label{app:mean_Delta_chi_K}
%%%%%%%%%%%%%%%%%%%%%%%%%%%%%%%%%%%%%%%%%%%%%%%%%%%%%%%%%%%%%%%%
%%%%%%%%%%%%%%%%%%%%%%%%%%%%%%%%%%%%%%%%%%%%%%%%%%%%%%%%%%%%%%%%

Let us compute the mean comoving distance~$\mean{\Delta \chi_{\rm K}}$ spent inside a Kottler hole. First note that, as already mentioned in Footnote~\ref{footnote:impact_parameter}, since the size of the holes is small compared to cosmological scales, we can reasonably consider
\begin{equation}
\Delta \chi_{\rm K} \approx f_K(\Delta \chi_{\rm K}) .
\end{equation}
Moreover, if a hole is crossed with (comoving areal) impact parameter~$B$, and neglecting light deflection, we have
\begin{equation}
f_K(\Delta \chi_{\rm K}) \approx 2 \sqrt{R_{\rm h}^2 - B^2},
\end{equation}
thus
\begin{equation}
\mean{\Delta \chi_{\rm K}}
\approx 
2\mean{\sqrt{R_{\rm h}^2 - B^2}}
=
2\int_{R_{\rm min}}^{R_{\rm max}} \dd R_{\rm h} \, p(R_{\rm h}) 
\int_{R_{\rm gal}}^{R_{\rm h}} \dd B \, p(B)\sqrt{R_{\rm h}^2 - B^2},
\end{equation}
where $R_{\rm min} \define R_{\rm h}(M_{\rm min}) = (3 M_{\rm min}/4\pi\rho_0)^{1/3}$, idem for $R_{\rm max}$. The integral over $B$ is easily calculated, and we finally obtain
\begin{align}
\mean{\Delta \chi_{\rm K}} 
&\approx \frac{4}{3} \int_{R_{\rm min}}^{R_{\rm max}} \dd R_{\rm h} \, p(R_{\rm h}) \, \sqrt{R_{\rm h}^2-R_{\rm gal}^2}
\\
&= \int_{M_{\rm min}}^{M_{\rm max}} \dd M p(M)
	\sqrt{R_{\rm h}^2(M)-R_{\rm gal}^2(M)} \\
&= \frac{4}{3} \pa{ \frac{3}{4\pi\rho_0} }^{1/3}
	\sqrt{ 1 - \pa{\frac{\rho_0}{\rho_{\rm gal}}}^{2/3} }
\int_{M_{\rm min}}^{M_{\rm max}} \dd M \, p(M) \, M^{1/3} \\
&\approx \frac{4}{3} \pa{ \frac{3}{4\pi\rho_0} }^{1/3}
\int_{M_{\rm min}}^{M_{\rm max}} \dd M \, p(M) \, M^{1/3} .
\end{align}

%%%%%%%%%%%%%%%%%%%%%%%%%%%%%%%%%%%%%%%%%%%%%%%%%%%%%%%%%%%%%%%%
%%%%%%%%%%%%%%%%%%%%%%%%%%%%%%%%%%%%%%%%%%%%%%%%%%%%%%%%%%%%%%%%

\bibliographystyle{JHEP}
\bibliography{bibliography}

%%%%%%%%%%%%%%%%%%%%%%%%%%%%%%%%%%%%%%%%%%%%%%%%%%%%%%%%%%%%%%%%
%%%%%%%%%%%%%%%%%%%%%%%%%%%%%%%%%%%%%%%%%%%%%%%%%%%%%%%%%%%%%%%%

\end{document}